\documentclass[12pt,a4paper,fleqn]{article}
\usepackage[T1]{fontenc}
\usepackage{amsmath}
\usepackage{color}
\usepackage{amssymb} 
\usepackage{bbm} 
\usepackage[small]{caption2} 
\usepackage{graphicx} 
\usepackage{mathrsfs}	
\usepackage[small,loose]{subfigure}  
\usepackage{cite} 
\advance \headheight by 3.0truept       

\newcommand{\eVdist}{\kern-0.06em}

\newcommand{\kev}{\:\text{ke\eVdist V}}
\newcommand{\kevee}{\:\text{ke\eVdist Vee}}

\newcommand{\gev}{\:\text{Ge\eVdist V}}
\newcommand{\tev}{\:\text{Te\eVdist V}}

\newcommand{\cm}{\,\text{cm}}
\newcommand{\sla}{\,\text{--}\,}

\allowdisplaybreaks

\begin{document}
\thispagestyle{empty}
\begin{titlepage}

\begin{flushright}
TUM-HEP 731/09
\end{flushright}

\vspace*{1.0cm}

\begin{center}
\Huge\textbf{Improved Constraints on Inelastic Dark Matter}
\end{center}
\vspace{1cm}
 \center{
\textbf{
Kai~Schmidt-Hoberg\footnote[1]{Email: \texttt{kschmidt@ph.tum.de}} and
Martin~Wolfgang~Winkler\footnote[2]{Email: \texttt{mwinkler@ph.tum.de}}
}
}
\\[5mm]
\begin{center}
\textit{\small
Physik-Department T30, Technische Universit\"at M\"unchen, \\
James-Franck-Stra\ss e, 85748 Garching, Germany
}
\end{center}

\vspace{1cm}

\begin{abstract}
We perform an extensive study of the DAMA annual modulation data in the context of inelastic dark matter.
We find that inelastic dark matter with mass $m_\chi\gtrsim 15\gev$ is excluded at the 95\% confidence level 
by the combination of DAMA spectral information and results from other direct 
detection experiments. However, at smaller $m_\chi$, inelastic dark matter constitutes a possible solution
to the DAMA puzzle.
\end{abstract}

\end{titlepage}

\newpage

\tableofcontents

\section{Introduction}

The DAMA collaboration has recently published the combined results from the long-term runs of the DAMA/NaI and DAMA/LIBRA
apparatus. Both experiments observe an annual modulation in their nuclear recoil rates with a combined statistical 
significance exceeding $8 \sigma$ \cite{Bernabei:2008yi}. 
A possible origin of this signal is weakly interacting massive particles (WIMPs) scattering off of the target nuclei in the
detector, where the modulation of the rate is caused by the motion of the earth around the sun.
The amplitude, period and phase of the signal are all consistent with such
scatterings.

This interpretation is challenged, however, by the absence of a signal in other dark matter direct detection experiments like 
CDMS~\cite{Ahmed:2008eu} and XENON~\cite{Angle:2007uj}. Especially in the standard scenario 
where WIMPs scatter elastically and spin-independently off of the 
target nucleus, the DAMA preferred region of WIMP cross sections is practically ruled out by the combination of 
other experiments~\cite{Fairbairn:2008gz,Savage:2008er}.

In order to reconcile the DAMA signal with the other direct detection experiments, a number of non-standard dark matter
candidates have been proposed. 
These include mirror dark matter~\cite{Foot:2003iv,Foot:2008nw}, dipolar dark matter~\cite{Masso:2009mu},
hidden charged dark matter~\cite{Feng:2009mn} and composite dark matter~\cite{Khlopov:2008zza}
as well as particles which couple spin-dependently~\cite{Savage:2004fn} or preferentially to electrons~\cite{Bernabei:2007gr,Kopp:2009et}.

In the present article we focus on inelastic dark matter~\cite{TuckerSmith:2001hy} which is one of the simplest 
possible solutions to the DAMA puzzle. Concrete realizations of inelastic dark matter in particle physics 
can e.g.~be found in~\cite{ArkaniHamed:2000bq,TuckerSmith:2001hy,Cui:2009xq,Arina:2009um}.
In this scenario the WIMP scatters from a ground state $\chi$ into the slightly heavier 
state $\chi'$.
If the mass splitting $\delta = m_{\chi'}-m_{\chi}$ is of the same order 
as the kinetic energy of WIMPs in the galactic halo, only the WIMPs in the tail of the velocity distribution 
are energetic enough to scatter off of a nucleus. 
In this case the the fraction of WIMPs which can overcome the 
kinematical barrier depends very strongly on the velocity of the earth, 
and the annual modulation of the WIMP scattering signal is significantly enhanced.
This raises the sensitivity of DAMA relative to other experiments and 
reduces the tension between them. Earlier studies indeed found parameter regions consistent with DAMA and all 
other experiments for WIMP masses $m_\chi\gtrsim 50\gev$ and $\delta = \mathcal{O}(100\kev)$ 
\cite{Chang:2008gd,Cui:2009xq,MarchRussell:2008dy}.

In this article we improve on those earlier studies. 
In particular we take into account the energy resolutions of the DAMA/LIBRA and DAMA/NaI apparatus and include higher 
energy bins which are typically ignored.
We also use more appropriate analysis techniques than some of the earlier works.
Furthermore we extend the analysis to small WIMP masses and find a region at $m_\chi \sim \mathcal{O}(10 \gev)$ 
where the DAMA signal can be explained through 
channeled scattering events off of iodine. Another allowed region at $m_\chi \sim 5\gev$ -- where the modulation is due to channeled 
sodium events -- opens up if one artificially decreases the detector resolutions of DAMA/LIBRA and DAMA/NaI.

We then confront the DAMA allowed regions with the null results of various other direct detection experiments. 
In contrast to previous studies we find that under standard astrophysical assumptions the whole high mass region
is ruled out at the 95\% confidence level (CL) by CRESST~II. 
For this statement to hold it is important that we do not combine the CRESST~II commissioning 
run~\cite{Angloher:2008jj} with an older test run~\cite{Angloher:2004tr}, as will be discussed in more detail later. 

The low mass regions corresponding to channeled scattering events at DAMA are partly probed by low-threshold experiments. 
However, here we still find regions at $m_\chi\lesssim 15\gev$ which are compatible with DAMA as well as all other direct detection experiments.

The outline of this paper is as follows: in the next section we review relevant aspects for 
the direct detection of inelastic dark matter. 
In Section~\ref{sec:experiments} we give an overview over the direct detection experiments we consider in this study 
and describe our analysis procedure. 
In Section~\ref{sec:results} we present the results of this analysis which we compare with results from 
previous studies in Section~\ref{sec:comparisonwithother}. Finally Section~\ref{sec:conclusions} contains our conclusions.

\section{Direct Detection of Inelastic Dark Matter}
\label{sec:overview}

Direct detection experiments aim to measure the deposited energy of a WIMP dark matter particle
when it interacts with a nucleus in the detector.
If such a dark matter particle $\chi$ scatters off of nuclei inelastically by making a transition to a 
slightly heavier state $\chi'$, the minimum incident velocity to transfer the energy $E_R$ to a recoiling 
nucleus is \cite{TuckerSmith:2001hy}
\begin{equation}
\label{eqn:idmvelo}
v_\text{min} = \frac{1}{\sqrt{2 m_N E_R}}\left( \frac{m_N E_R}{\mu} + \delta \right) \;,
\end{equation}  
where $\delta$ is the mass splitting between $\chi$ and $\chi^\prime$, $m_N$ is the mass of the target nucleus
and $\mu=(m_\chi m_N)/(m_\chi+ m_N)$ is the reduced mass of the WIMP-nucleus system. The altered kinematics of the scattering
process relative to elastic scattering can affect the detection rates significantly.
We will discuss these effects in more detail after a short 
reminder of the relevant formulae 
for the calculation of reaction rates
and a brief discussion of the relevant input parameters.
We will also specify the impact of detector effects on reaction rates.

\subsection{Reaction Rates}
  
The differential event rate for coherent spin-independent
WIMP nucleus scattering as a function of the recoil energy $E_R$ can be written as (see e.g.~\cite{Lewin:1995rx})
\begin{equation}\label{eq:DifferentialEvents}
\frac{\text{d}R}{\text{d}E_R}(E_R,t)=M_\text{tar} \frac{\rho_\chi}{2\, m_\chi \mu^2}\, 
\frac{(f_p Z+ f_n(A-Z))^2}{f_n^2}\, \sigma_n\;F^2(E_R)
\int\limits_{v_{\text{min}}}^\infty\!\!\!\text{d}^3 v\; \frac{f_\text{local}(\vec{v},t)}{v} \;,
\end{equation}
with $M_\text{tar}$ the mass of the target detector, $m_\chi$ the WIMP mass and $\rho_\chi$ the local WIMP density. 
Furthermore $f_{n,p}$ are the 
effective coherent couplings to the neutron and proton respectively, while  
$A$ and $Z$ denote the nucleon and proton numbers of the target nucleus and $\sigma_n$ the overall effective 
WIMP-neutron cross section at zero momentum in the elastic limit. The nuclear form factor $F(E_R)$ describes the loss of coherence as the 
momentum transfer deviates from zero. Finally, the integral takes into account the effect of 
the local WIMP velocity distribution $f_\text{local}$.
Both, the input from particle physics as well as the astrophysical input together with the corresponding uncertainties will be addressed more closely below.

\subsection{Input Parameters}\label{sec:generalInput}

To keep our analysis as general as possible, we keep the WIMP mass $m_\chi$, the mass splitting $\delta$ and 
the cross section $\sigma_n$ as free parameters. These parameters may be predicted by the particle physics model 
under consideration.

\subsubsection{Input from Particle Physics}

For this study we set the ratio of the effective WIMP-proton and WIMP-neutron couplings to one, $f_p/f_n=1$. 
This ratio could be different in a specific model. However, a ratio $f_p/f_n \neq 1$ can easily be corrected for.  
One just has to rescale the considered constraint on $\sigma_n$ by a factor 
\begin{equation}
\left[
1+\left(\frac{f_p}{f_n}-1 \right)\,\frac{Z}{A}\right]^{-2} \;.
\end{equation}
As all nuclei have roughly the same ratio of protons to neutrons, a modification of $f_p/f_n$ does virtually 
not change the experimental constraints relative to each other.

In our analysis we use the Fourier-Bessel form factors where available, otherwise the Woods-Saxon form 
factors\footnote{Fourier-Bessel form factors are available for oxygen, aluminum, silicon and germanium, 
Woods-Saxon form factors are used for sodium, iodine, xenon and tungsten.}. 
Both can be calculated from the parameters tabulated in~\cite{DeJager:1987qc,Fricke}. Note that Fourier-Bessel 
and Woods-Saxon form factors are more accurate than the commonly used Helm form factors (see e.g.~\cite{Duda:2006uk}). 
Especially for heavy elements the Woods-Saxon form factors can deviate from the Helm form factors substantially. For example in the case 
of a tungsten target as used by CRESST~II the Helm form factor would lead to an overestimation of the reaction rate by roughly $10\%$.

\subsubsection{Astrophysical Input}

The two main astrophysical inputs used in this study are the local dark matter density 
$\rho_\chi$ and the dark matter velocity distribution seen by an earth bound detector.
For our analysis, we set $\rho_\chi=0.3\gev\cm^{-3}$. Note that a change of $\rho_\chi$ would just lead to an overall shift of 
all experimental constraints on $\sigma_n$.

The galactic dark matter halo is expected to have a rather smooth velocity distribution. The Standard Halo Model, which we adopt 
throughout this study, assumes a Maxwell-Boltzmann distribution in the galactic rest frame, which is truncated at some escape 
velocity $v_\text{esc}$~\cite{Drukier:1986tm},
\begin{equation}
f_{\text{gal}}(\vec{v})=\frac{1}{(\pi\, v_0^2)^{3/2}}\,\, e^{-\vec{v}^2/v_0^2}\;\Theta(v_{\text{esc}}-|\vec{v}|)\,.
\end{equation}
The local galactic escape velocity was determined to be in the interval
$498\,\text{km/s} < v_{\text{esc}} < 608\,\text{km/s}$ at 90\% CL by the RAVE survey~\cite{Smith:2006ym}. We will use the median 
$v_{\text{esc}}=544\,\text{km/s}$. The velocity dispersion is set to its standard value $v_0 = 220\,\text{km/s}$.

The dark matter velocity distribution in the rest frame of the earth, $f_{\text{local}}$, can be obtained 
from the galactic distribution by a Galilean boost, 
\begin{equation}
f_\text{local}(\vec{v},t)=f_{\text{gal}}(\vec{v}+\vec{v}_{\text{earth}}(t))\,,
\end{equation}
where the velocity of the earth relative to the galactic halo can be parameterized as
\begin{equation}
\vec{v}_{\text{earth}}(t)= \vec{v}_{\text{sun}} + \vec{v}_{\text{ann}}(t)\,.
\end{equation}
Using the convention for galactic coordinates as in~\cite{Gelmini:2000dm} we have
\begin{equation}
\vec{v}_{\text{sun}}\simeq \begin{pmatrix}0\\220\\0 \end{pmatrix}\,\text{km/s}\,+ \begin{pmatrix}10\\5.2\\7.2 
\end{pmatrix}\,\text{km/s}
\end{equation}
describing the motion of the solar system~\cite{Dehnen:1997cq,Binney:2008aa} and
\begin{equation}
\vec{v}_{\text{ann}}(t)\simeq 29.8\,\text{km/s}\left[ \begin{pmatrix}0.9931\\0.1170\\-0.01032 \end{pmatrix}\,
\cos{(2\pi(t-t_1))}+ \begin{pmatrix}-0.0670\\0.4927\\-0.8678 \end{pmatrix}\,\sin{(2\pi(t-t_1))}\right] 
\end{equation}
corresponding to the time-dependent velocity of the earth relative to the sun~\cite{Lewin:1995rx,Gelmini:2000dm}.
Here $t$ is measured in years and $t_1=0.219$ corresponds to the Spring equinox on March~21.
The maximal and minimal velocities of the earth with respect to the galactic rest frame are reached on June~2 and December~2
respectively.

\subsection{Annual Modulation and Inelastic Dark Matter}
 
The motion of the earth around the sun leads to an annual modulation of the WIMP velocity distribution
with respect to the earth. Due to this annual modulation, the differential rate of WIMP scattering 
has a constant and a nearly sinusoidal time-dependent contribution,
\begin{equation}\label{eq:generalmodulation}
\frac{\text{d} R}{\text{d} E_R}(E_R,t)=S_0 + S_m \cos{(2\pi (t-t_0))}\; ,
\end{equation}
with $t_0$ corresponding to June 2nd, where the relative earth-WIMP velocity is maximal. 
Note that for non-standard velocity distributions, which have additional features, the time dependence could in principle have a more complicated form.
The modulation amplitude of the signal $S_m$ is given by
\begin{equation}
S_m(E_R)=\frac{1}{2} \left( \frac{\text{d}R}{\text{d}E_R}(E_R, \text{June 2})-\frac{\text{d}R}{\text{d}E_R}(E_R, \text{Dec 2})\right)\,.
\end{equation}
Since DAMA, unlike other direct detection experiments, does not attempt to suppress backgrounds completely, it 
can only identify the modulated part of the signal. Taken by itself, the DAMA signal is consistent with the standard 
scenario of elastic WIMP scattering, even though the modulation $S_m/S_0$ is expected to be only at the level of a few percent in this case.
This interpretation is, however, inconsistent with several other direct detection experiments and if
the DAMA signal really is due to WIMP-nucleus scatterings, other dark matter interpretations have to be considered.

One proposal to reconcile the DAMA signal with the null results from other experiments
is inelastic dark matter. The inelasticity of the scattering drastically changes the kinematics of the process, c.f.~\eqref{eqn:idmvelo}. 
In particular the higher kinematical barrier for inelastic processes allows only WIMPs in the tail of the velocity 
distribution to scatter. This leads to a strong enhancement of the annual modulation of the scattering rate --
in the extreme case that the kinetic energies of WIMPs needed to scatter inelastically are reached only during summer, 
the modulation can even be maximal\footnote{See e.g.~the left panel of Figure~\ref{fig:dama100gev} for the typical annual modulation of the recoil rates.}.
This effect raises the sensitivity of DAMA compared to other direct detection experiments and ameliorates the tension
between them.   
Furthermore, as the minimal velocity~\eqref{eqn:idmvelo} typically decreases at higher nucleus mass, 
experiments with heavy target nuclei can detect a larger fraction of the WIMPs passing through the detector. 
This implies e.g.~that the sensitivity of CDMS is reduced relative to DAMA, since both germanium and silicon are
considerably lighter than iodine. On the other hand xenon is similar in mass and hence the XENON10 experiment should 
be sensitive to inelastic recoils. But since XENON10 took data only during winter where the reaction rates are smallest, 
the resulting limits can still be evaded. Nevertheless the other experiments still very strongly constrain 
the inelastic dark matter interpretation, with especially strong constraints arising from the CRESST~II experiment, 
which has tungsten as the heaviest target nucleus.

\subsection{Detector Effects}

In a real experiment there are additional effects which have to be taken into account to understand a possible signal. 
First of all, only a fraction of the energy of the recoiling nucleus will be transferred to the channel which
is observed, typically as ionization or scintillation in the detector. This fraction is known as the 
quenching factor $Q$ which relates the recoil energy $E_R$ to the observed energy $E^\prime$ through 
\begin{equation}\label{eq:quenching}
E^\prime=Q E_R \;.
\end{equation}
The observed energy $E^\prime$ is often referred to as electron equivalent energy and measured in keVee.
The quenching factor mainly depends on the target material of the detector. 
Some experiments calibrate their energy scales such that the quenching factor is effectively one. 
For experiments with $Q\neq1$ one can apply~\eqref{eq:quenching} to obtain the measured events in terms of the recoil energy.

Assuming an ideal experimental setup the event rate 
would simply be given by~\eqref{eq:DifferentialEvents}. However, a realistic detector has only a finite energy resolution, i.e.~for a 
recoiling nucleus with energy $E_R$ the measured energy $E^\prime$ follows a probability distribution which is peaked at $Q\,E_R$. 
Typically one approximates the distribution by a Gaussian with an energy dependent standard deviation $\sigma(E^\prime)$. To obtain 
the predicted differential reaction rate corrected by the finite energy resolution of the detector, one must convolute 
$\text{d}R / \text{d}E_R$ from~\eqref{eq:DifferentialEvents} with this Gaussian,
\begin{equation}\label{eq:resolutionquenching}
\frac{\text{d}R}{\text{d}E_R}_\text{\tiny{Corrected}}\hspace{-1.15cm}(E_R,t)= \frac{1}{\sqrt{2\pi}} 
\int\limits_0^{\infty} \text{d}E_R' \, \frac{\text{d}R}{\text{d}E_R'}(E_R',t)\;\frac{1}{\sigma(E_R')} \;\text{exp}\left(-{\frac{(E_R-E_R')^2}
{2\sigma^2(E_R')}}\right)  \;,
\end{equation}
where $\sigma(E_R)$ may be gained from $\sigma(E^\prime)$ by use of the quenching factor.

Finally there is an energy dependent efficiency $\xi$, which takes into account the loss of signal, e.g.~due to data cuts 
which are needed to reduce background. To determine the number of recoils which are expected to be seen at a 
given experiment between recoil energies $E_1$ and $E_2$ in a time interval $t_1 \sla t_2$, we have to integrate over the 
resolution corrected rate multiplied by the efficiency,
\begin{equation}\label{eq:Events}
N_{E_1-E_2}=\int\limits_{t_1}^{t_2} \text{d}t \int\limits_{E_1}^{E_2} \text{d}E_R \,\xi(E_R)\, 
\frac{\text{d}R}{\text{d}E_R}_\text{\tiny{Corrected}}\hspace{-1.15cm}(E_R,t) \;.
\end{equation}
For detectors with multiple elements/isotopes, the contributions for each element/isotope have to be added.

\section{Direct Detection Experiments}
\label{sec:experiments}

In this section we discuss key properties and results from the DAMA experiment as well
as from various null searches. We restrict ourselves to experiments which provide
sufficient data to independently generate constraints.

\subsection{DAMA}

The DAMA collaboration has recently published the observed modulation amplitudes in 36 energy bins over $2\sla 20\kevee$, where the data from 
DAMA/LIBRA and DAMA/NaI were combined. We will shortly describe how to calculate the modulation amplitude expected at DAMA and then 
outline the statistical procedures used to determine the confidence regions for DAMA.

\subsubsection{Modulation Amplitude at DAMA}

DAMA's target consists of highly radiopure NaI(Tl) crystals.
In principle both components -- sodium and iodine -- may contribute to the observed signal. Their respective quenching factors 
were measured to be $Q_\mathrm{Na} = 0.3$ and $Q_\mathrm{I}=0.09$~\cite{Bernabei:1996vj} respectively. 

While typically only this small fraction of the recoil energy goes into scintillation which can be measured by DAMA, some 
nuclei that recoil
along the characteristic axes or planes of the crystal structure may transfer their energy primarily to electrons resulting 
in $Q \sim 1$~\cite{Bernabei:2007hw}. 
This effect is known as channeling.
DAMA has used simulations to determine the energy-dependent fractions
of channeled events off of sodium and iodine. These fractions can be extracted from
Figure~4 in~\cite{Bernabei:2007hw}. A good analytical approximation is given by
\begin{equation} \label{eqn:fNa}
f_\mathrm{Na}(E_R) = 0.63^{E_R^{0.49}\left(1+\frac{E_R}{1+E_R}\right)} \quad\quad \text{and} \quad\quad  
f_\mathrm{I}(E_R) = 0.66^{E_R^{0.45}\left(1+\frac{E_R}{1+E_R}\right)}
\end{equation}
respectively, where the recoil energy $E_R$ is understood in units of keV. In order to be able to compare predicted and measured rates,
it is useful to calculate the predicted spectrum in terms of the measured energy. Taking into account 
quenching and channeling and adding contributions from sodium and iodine, one finds
\begin{align}
\frac{\text{d}R}{\text{d}E^\prime}(E^\prime,t)&= 0.847 \left[ 
 f_\mathrm{I}(E^\prime)\;\, \frac{\text{d}R_\mathrm{I}}{\text{d}E_R}(E^\prime,t) 
+\frac{1-f_\mathrm{I}\!\left(\frac{E^\prime}{Q_\mathrm{I}}\right)}{Q_\mathrm{I}}\;\;\frac{\text{d}R_\mathrm{I}}{\text{d}E_R}\left(\frac{E^\prime}{Q_\mathrm{I}},t\right)
\right] \nonumber\\
&+ 0.153 \left[ 
 f_\mathrm{Na}(E^\prime)\; \frac{\text{d}R_\mathrm{Na}}{\text{d}E_R}(E^\prime,t) 
+\frac{1-f_\mathrm{Na}\!\left(\frac{E^\prime}{Q_\mathrm{Na}}\right)}{Q_\mathrm{Na}}\;\;\frac{\text{d}R_\mathrm{Na}}{\text{d}E_R}\left(\frac{E^\prime}{Q_\mathrm{Na}},t\right)
\right],
\end{align}
where $\text{d}R_\mathrm{I}/\text{d}E_R$  and $\text{d}R_\mathrm{Na}/\text{d}E_R$ may be gained from~\eqref{eq:DifferentialEvents}, 
while 0.847 and 0.153 denote the mass fractions of iodine and sodium in NaI respectively. The event rate has then to be 
corrected by the detector resolution according to~\eqref{eq:resolutionquenching}. For DAMA/LIBRA $\sigma(E^\prime)$ was 
determined to be\cite{Bernabei:2008yh}
\begin{equation} \label{eqn:DAMAER}
  \sigma_\text{DAMA/LIBRA}(E^\prime) = (0.448\kevee) \sqrt{\frac{E^\prime}{\mathrm{keVee}}}
                     + 0.0091 E^\prime\;.
\end{equation}
The resolution of DAMA/NaI is slightly worse, it can be obtained by a simple fit to Figure~13 in~\cite{Bernabei:1998rv},
\begin{equation} \label{eqn:DAMANaIER}
  \sigma_\text{DAMA/NaI}(E^\prime) = (0.76\kevee) \sqrt{\frac{E^\prime}{\mathrm{keVee}}}
                     - 0.024 E^\prime .
\end{equation}
The different energy resolutions of the two experiments force us to calculate the resolution corrected spectra of 
DAMA/LIBRA and DAMA/NaI separately. We then combine them weighted by their relative exposures which are 65\% and 35\% respectively,
\begin{equation}
 \frac{\text{d}R}{\text{d}E^\prime}_\text{\tiny{Corrected}}\hspace{-1.15cm}(E^\prime,t)\,
= 0.65\, \left( \frac{\text{d}R}{\text{d}E^\prime}_\text{\tiny{Corrected}}\hspace{-1.15cm}(E^\prime,t)\,\right)_\text{DAMA/LIBRA}
+ 0.35\, \left( \frac{\text{d}R}{\text{d}E^\prime}_\text{\tiny{Corrected}}\hspace{-1.15cm}(E^\prime,t)\,\right)_\text{DAMA/NaI}\;.
\end{equation}
Finally the predicted modulation amplitude averaged over an energy bin reads  
\begin{equation}
 S^\text{bin}_\text{pred}=\frac{1}{2\,(E^\prime_2-E^\prime_1)} \int\limits_{E^\prime_1}^{E^\prime_2} \text{d}E^\prime
 \left( \frac{\text{d}R}{\text{d}E^\prime}_\text{\tiny{Corrected}}\hspace{-1.15cm}(E^\prime, \text{June 2}) -
 \frac{\text{d}R}{\text{d}E^\prime}_\text{\tiny{Corrected}}\hspace{-1.15cm}(E^\prime, \text{Dec 2})\right) \;.
\end{equation}

\subsubsection{Calculating Limits}\label{sec:damalimits}

To find regions in the parameter space of inelastic dark matter which are consistent with DAMA, we employ a 
$\chi^2$-goodness-of-fit metric,
\begin{equation}
 \chi^2=\sum\limits_{\text{bins}} \frac{\left( S^\text{bin}_\text{pred}-S^\text{bin}_\text{data}\right)^2}
{({\sigma}^\text{bin}_\text{data})^2}\,,
\end{equation}
where $S^\text{bin}_\text{data}$ stands for the measured modulation 
amplitude, while ${\sigma}^\text{bin}_\text{data}$ denotes the corresponding experimental error.
The modulation signal is consistent with zero above $10\kevee$, therefore it is not appropriate to simply fit to all 36 energy bins of DAMA. 
Fitting to a large number of bins, where the measured modulation amplitude fluctuates around zero with a relatively large error, 
would just dilute the power of the goodness of fit metric. In~\cite{Chang:2008gd,Cui:2009xq,MarchRussell:2008dy} this problem was circumvented by just 
ignoring the higher energy bins of DAMA. However, we 
follow~\cite{Savage:2008er} where all bins above $10\kevee$ are combined into a single bin. 
The inclusion of this additional bin makes sure that scenarios in which the modulation amplitude is 
non-zero above $10\kevee$ are strongly constrained.

In total we use 17 energy bins for the goodness-of-fit test, cf.~Table~\ref{tab:DAMABinned}.
As we are fitting to 17 energy bins, the $90\%$ and $99\%$-confidence levels are characterized by 
$\chi^2<24.8$ and $\chi^2<33.4$ respectively.
\begin{table}  
\centering
  \begin{tabular}{|c c | c c|}
    \hline\hline
    Energy  & $S^\text{bin}_\text{data}$  & Energy  & $S^\text{bin}_\text{data}$ \\\empty
    [keVee] & [cpd/kg/keV]  & [keVee] & [cpd/kg/keV]  \\
    \hline
    2.0\sla2.5  &  0.0162 $\pm$ 0.0048 &  6.5\sla7.0  &  0.0012 $\pm$ 0.0036  \\
    2.5\sla3.0  &  0.0287 $\pm$ 0.0054 &  7.0\sla7.5  & -0.0002 $\pm$ 0.0036 \\
    3.0\sla3.5  &  0.0250 $\pm$ 0.0055 &  7.5\sla8.0  &  0.0004 $\pm$ 0.0036  \\
    3.5\sla4.0  &  0.0141 $\pm$ 0.0050 &   8.0\sla8.5  & -0.0014 $\pm$ 0.0037 \\
    4.0\sla4.5  &  0.0100 $\pm$ 0.0045 &   8.5\sla9.0  &  0.0039 $\pm$ 0.0037 \\
    4.5\sla5.0  &  0.0118 $\pm$ 0.0040 &  9.0\sla9.5  & -0.0033 $\pm$ 0.0037  \\
    5.0\sla5.5  &  0.0039 $\pm$ 0.0040 &  9.5\sla10.0 & -0.0070 $\pm$ 0.0038  \\
    5.5\sla6.0  &  0.0030 $\pm$ 0.0039 &    & \\ \cline{3-4}
    6.0\sla6.5  &  0.0060 $\pm$ 0.0038 &  10.0\sla20.0 & 0.0010  $\pm$ 0.0008  \\
       \hline\hline
  \end{tabular}
  \caption{DAMA modulation amplitudes (and errors) in the lowest sixteen energy bins
           between 2 and $10\kevee$, extracted from Figure~9 in~\cite{Bernabei:2008yi}. 
           Bins above $10\kevee$ are combined into a single bin.}
\label{tab:DAMABinned}
\end{table}

The DAMA collaboration has also published the absolute rate $S_0$ (cf.~\eqref{eq:generalmodulation}) measured by the DAMA/LIBRA apparatus 
in the range $0.75\sla 10\kevee$ (cf.~Figure~1 in~\cite{Bernabei:2008yi}). As there is only limited background suppression, a large 
fraction of the measured rate may be due to background events. However, this data set can still be used to derive conservative exclusion limits 
by imposing that the predicted absolute rate may at least not exceed the measured rate\footnote{One does not have to include a statistical error 
in this calculation as the number of events observed in each bin is extremely large.}. For this constraint we use only the energy bins above 
$2\kevee$ as the behavior of the detector below its threshold may not be well understood.

\subsection{Null Searches}

Here we list some of the experiments which give the most stringent bounds on the dark matter
interpretation of the DAMA signal. Key parameters of the experiments we use
can be found in Table~\ref{tab:ExpParam}. While most of these experiments observe some recoils, these
recoils may be due to background events. In this work we do not attempt to subtract the background, 
hence we obtain conservative constraints.

\begin{table*}
  \begin{tabular}{l|cccccc}
  \hline\hline
  Experiment   & XENON10 & ZEPLIN~III & CDMS~II & CDMS-SUF   \\   
  \hline
  Target
    & Xe
    & Xe
    & Ge
    & Ge \\
  Exp. [kg-day]
    & 316.4
    & 126.7
    & 397.8
    & 65.8 \\  
  Range [keVee]
    & $6.1\sla 26.9$
    & $2\sla 16$
    & $10\sla 100$
    & $5\sla 100$ \\  
  Q
    & 1 
    & \eqref{eqn:ZEPLINQ}
    & 1 
    & 1  \\
  Efficiency [Eq.]
    &  \eqref{eqn:XENON10eff}
    &  \eqref{eqn:ZEPLINeff}
    &  \eqref{eqn:CDMSIIeff}
    &  \eqref{eqn:CDMSIeff}\\
  Resolution [Eq.]
    &  \eqref{eqn:XENON10ER}
    &  \eqref{eqn:XENON10ER}  
    &  \eqref{eqn:CDMSIIER}
    &  \eqref{eqn:CDMSIIER}                 \\
  $t_1[y]$
    & 0.763
    & 0.158
    & $\sim$ 0.80
    & 0.995 \\  
  $t_2[y]$
    & 1.122
    & 0.382
    & $\sim$ 1.55
    & 1.248 \\
  Constraint
    & maximum gap
    & maximum gap
    & maximum gap
    & maximum gap \\  
  \hline\hline
  Experiment & CRESST~I & CRESST~II & CoGeNT & TEXONO \\
\hline
  Target
    & Al, O
    & W
       & Ge
    & Ge       \\
  Exp. [kg-day]
    & 0.80, 0.71
    & 30.6
    & 8.38
    & 0.338 \\  
  Range [keVee]
    & $0.6\sla 20$
    & $10\sla 100$
    & $0.34\sla 4.1$
    & $0.2\sla 7.1$  \\  
  Q
    & 1 
    & 1 
    & 0.2
    & 0.2   \\
  Efficiency [Eq.]
    & 1
    & 0.9
    &  \eqref{eqn:CoGeNTeff}
    &  \eqref{eqn:TEXONOeff}\\
  Resolution [Eq.]
    &  \eqref{eqn:CRESSTIER}
    &  \eqref{eqn:CRESSTIIER}
    &  \eqref{eqn:CoGeNTER}
    &  \eqref{eqn:TEXONOER}         \\
  $t_1[y]$
    & $\sim$ 0.75
    & 0.234
    & $\sim$ 0.17
    & 0.040   \\  
  $t_2[y]$
    & $\sim$ 0.75
    & 0.558
    & $\sim$ 0.25
    & 0.100    \\
  Constraint
    & binned Poisson
    & maximum gap
    & binned Poisson
    & binned Poisson \\  
  \hline\hline
  \end{tabular} 
  \caption{
    Parameters of direct dark matter searches for 
    XENON10~\cite{Angle:2007uj}, ZEPLIN~III~\cite{Lebedenko:2008gb}, CDMS-SUF~\cite{Akerib:2003px}, CDMS~II~\cite{Ahmed:2008eu}, 
    CRESST~I~\cite{Angloher:2002in}, CRESST~II~\cite{Angloher:2008jj}, CoGeNT~\cite{Aalseth:2008rx} and TEXONO~\cite{Lin:2007ka}.
    Note that for CoGeNT and Texono we only use the energy bins below $2\kevee$ in our analysis (see text).
  }
  \label{tab:ExpParam}
\end{table*}

For unbinned data, constraints are generated
using S.~Yellin's maximum gap method \cite{Yellin:2002xd}, which
examines the likelihood of observing the gaps in energy between
observed events.  This unbinned method generally provides a stronger
constraint than binned methods for the case of an unknown background.
Unfortunately experimental data is often available in binned form only.
In this case we use the 'binned Poisson' technique, as described in~\cite{Green:2001xy,Savage:2008er}.

\subsubsection{XENON10}

The XENON10 detector acquired 316.4~kg-days of data between October 6, 2006 and
February 14, 2007 and observed 10 candidate events~\cite{Angle:2007uj}.
Here energies are calibrated to recoil energies, implying $Q=1$.  
This calibration depends upon the relative scintillation efficiency $\mathcal{L}_{\mathrm{eff}}$
which was assumed to be constant, $\mathcal{L}_{\mathrm{eff}}=0.19$~\cite{Angle:2007uj}. However, new 
measurements~\cite{Sorensen:2008ec,Aprile:2008rc} indicate that the scintillation efficiency decreases to 
$\mathcal{L}_{\mathrm{eff}} \simeq 0.14$ for recoil energies $E_R\lesssim 10 \kev$ (see Figure~8 in~\cite{Aprile:2008rc}). 
We account for the lower $\mathcal{L}_{\mathrm{eff}}$ by rescaling the lower threshold of XENON10 and the lowest energy event 
by a factor $0.19/0.14$. All other events are located at larger recoil energies and are not affected. After our rescaling 
the energy range of XENON10 is $6.1\sla 26.9\kev$.

The energy dependent efficiency can be approximated by (c.f.~Table~1 in~\cite{Angle:2007uj})
\begin{equation} \label{eqn:XENON10eff}
\xi(E_R) = 0.46 \left(1 - \frac{E_R}{100\kev}\right)
\end{equation} 
while the energy resolution can be written as~\cite{Savage:2008er}
\begin{equation} \label{eqn:XENON10ER}
  \sigma(E_R) = (0.579\kev) \sqrt{\frac{E_R}{\mathrm{keV}}}
                     + 0.021 E_R \;.
\end{equation}

\subsubsection{ZEPLIN~III}

Between February 27 and May 20, 2008,  ZEPLIN~III  acquired 126.7 kg-days of data~\cite{Lebedenko:2008gb},
taking into account all the energy independent efficiencies. Seven events were detected in the observed energy range
$2\sla 16\kevee$, corresponding to recoil energies $10.7\sla 30.2\kev$, cf.~Figure~12 in~\cite{Lebedenko:2008gb}.
The quenching factor $Q$ has to be extracted from the conversion between the energy scales
in Figure~15, an analytic approximation is given by~\cite{MarchRussell:2008dy} 
\begin{equation}
\label{eqn:ZEPLINQ}
Q(E^\prime)=(0.142 E^\prime + 0.005) \, \text{Exp}(-0.305 E^{'0.564}) \;.
\end{equation}
The energy dependent efficiency is also shown in Figure~15 of~\cite{Lebedenko:2008gb}. 
In the considered window of recoil energies it is well approximated by
\begin{equation}
\label{eqn:ZEPLINeff}
\xi(E_R)=0.71 +\frac{15.23 \kev}{E_R}-\frac{197.3 \kev^2}{E_R^2} \;.
\end{equation}
Unfortunately there is no information available on the energy resolution of ZEPLIN~III. However, as XENON10 and ZEPLIN~III 
have a similar detector setup, we will employ~\eqref{eqn:XENON10eff} also for ZEPLIN~III.

\subsubsection{CDMS~II}
CDMS~II took data from October 2006 to July 2007, resulting in an exposure
of 397.8~kg-days on its germanium detectors with an energy threshold of 10~keV \cite{Ahmed:2008eu}.
No events were observed.
The quenching factor is $Q=1$.
The efficiency of observing nuclear recoils can be extracted from~Figure~2 in~\cite{Ahmed:2008eu} (arXiv version only), we use the approximation
\begin{equation} \label{eqn:CDMSIIeff}
  \xi(E_R) =
    \begin{cases}
      0.24 +\frac{3.39\kev}{E_R}-\frac{35.3 \kev^2}{E_R^2}
        & \textrm{for} \,\, 10\kev < E_R < 40\kev \; , \\
      0.30
        & \textrm{for} \,\, 40\kev < E_R < 100\kev \; .
    \end{cases}
\end{equation}
The energy resolution for the CDMS germanium detectors is given by~\cite{cdmspc}
\begin{equation} 
\label{eqn:CDMSIIER}
  \sigma(E_R) \simeq 0.2 \sqrt{E_R} \;.
\end{equation}

\subsubsection{CDMS-SUF}
Due to the low $5\kev$ threshold we also include the 65.8~kg-days germanium
data set from CDMS-SUF \cite{Akerib:2003px} in our analysis, which was taken from December 2001 to
April 2002.

The total efficiency can be approximated by \cite{Ahmed:2008eu}
\begin{equation} \label{eqn:CDMSIeff}
  \xi(E_R) = 0.80 \times 0.95 \times
    \begin{cases}
      0.10 + 0.30 \frac{(E_R - 5\kev)}{15\kev}
        & \textrm{for} \,\, 5\kev < E_R < 20\kev , \\
      0.40 + 0.10 \frac{(E_R - 20\kev)}{80\kev}
        & \textrm{for} \,\, 20\kev < E_R < 100\kev .
    \end{cases}
\end{equation}
In total there were 20 candidate events in the germanium detectors in the 
recoil energy range $5\sla100\kev$ which are however consistent with expected backgrounds. 
For the energy resolution we apply~\eqref{eqn:CDMSIIER} as for CDMS~II. We have also checked that the silicon data set from the same run gives 
no further constraints.

\subsubsection{CRESST~I}

CRESST~I took 1.51~kg-days of data in October 2000 \cite{Angloher:2002in}.  
The quenching factor is $Q=1$.
The data is binned from 0.6 to 20 keV, each bin covers an energy range of
0.2 keV. We use the events after cuts have been taken into account, cf.~Figure~9 \& 10 in \cite{Angloher:2002in}.
The energy resolution given by the CRESST collaboration is~\cite{Angloher:2002in}
\begin{equation} \label{eqn:CRESSTIER}
  \sigma(E_R) = \sqrt{(0.220\kev)^2 + (0.017 E_R)^2}\;.
\end{equation}

\subsubsection{CRESST~II}\label{sec:cresst2}

Two CaWO$_4$ crystal CRESST~II detector modules (Verena and Zora) took data between March~27 and July~23, 2007.
The total exposure of tungsten corresponds to $30.6$~kg-days in an energy range of $10\sla 100\kev$ with an acceptance of 0.9. Scattering off of 
calcium and oxygen can safely be neglected. Seven candidate events are shown in Figure~8 of~\cite{Angloher:2008jj}.
For the energy resolution we use\footnote{The energy resolution does not exceed $1 \mathrm{keV}$ 
in the energy range up to $100\kev$\cite{cresstpc}, a slight overestimation of the resolution will only make our 
limits marginally more conservative.}
\begin{equation} \label{eqn:CRESSTIIER}
  \sigma(E_R) \sim 1 \mathrm{keV} \;.
\end{equation}

There exists also a data set from an older run of CRESST~II with the prototype detector modules Daisy and Julia~\cite{Angloher:2004tr}. 
Some previous studies~\cite{MarchRussell:2008dy,Chang:2008gd} combined the commissioning run with this older test run. 
We emphasize that we refrain from doing this. The reason is that the CRESST~II detector was subject to significant modification 
between the runs, in particular a neutron shield as well as a muon veto were installed~\cite{Angloher:2008jj}. The absence of the 
neutron shield in the test run in combination with the poor light resolution of the Julia module may have led to a "leakage" 
of neutron-induced recoils into the region of tungsten events~\cite{Angloher:2004tr}. It may well be due to this particular 
background that the observed event rate in the test run was higher than in the commissioning run. Therefore we find it most 
appropriate to include only the commissioning run in our analysis, which leads to stronger constraints.

\subsubsection{CoGeNT}

CoGeNT took 8.38~kg-days of data in March 2008 above energies of $\sim 0.3\kevee$~\cite{Aalseth:2008rx,cogentpc}. 
The quenching factor is $Q=0.2$. 
We obtained a data sample from the CoGeNT collaboration containing the absolute number of events in very small bins of size $\sim 30$~eVee. 
One should, however, not use such small bins for the analysis, as gain-instabilities can shift the measured energy of events slightly
and introduce systematical errors~\cite{cogentpc}.
We therefore followed the procedure of the CoGeNT collaboration~\cite{Aalseth:2008rx,cogentpc} and combined four bins into one,
resulting in a bin width of $\sim 130\kevee$, c.f.~Table~\ref{tab:CoGeNT}. 
We ignored all bins above $E^\prime=2\kevee$ corresponding to $E_R=10\kev$ in order not to dilute the statistical power
of the binned Poisson method. Above this energy CDMS~II is very sensitive 
and CoGeNT cannot give additional limits.

We checked that the efficiency of CoGeNT is well approximated by the formula given in~\cite{Savage:2008er},
\begin{equation} 
\label{eqn:CoGeNTeff}
  \xi(E^\prime) = 0.66 - \frac{E^\prime}{50\kevee} \;.
\end{equation}
The energy resolution used by CoGeNT is~\cite{Savage:2008er}
\begin{equation} \label{eqn:CoGeNTER}
  \sigma(E^\prime) = \sqrt{(69.7\,\mathrm{eVee})^2
                           + (0.98\,\mathrm{eVee}) E^\prime} \;.
\end{equation}

\begin{table}
\centering
  \begin{tabular}{|cc|cc|}
\hline
   Energy  & Events  & Energy  & Events \\
    \hphantom{}[keVee] &         & [keVee] &        \\
\hline
    0.338\sla0.467 & 5703
     &  1.241\sla1.370 & 62 \\
    0.467\sla0.596 & 52
     &  1.370\sla1.499 & 38 \\
    0.596\sla0.725 & 33
     &  1.499\sla1.628 & 35 \\
    0.725\sla0.854 & 50
     &  1.628\sla1.757 & 31 \\
    0.854\sla0.983 & 45
     &  1.757\sla1.886 & 25 \\
    0.983\sla1.112 & 32
     &  1.886\sla2.015 & 19 \\
    1.112\sla1.241 & 50 & & \\
\hline
  \end{tabular} 
  \caption{
    CoGeNT binned data up to $2\kevee$.
  }
  \label{tab:CoGeNT}
\end{table}

\subsubsection{TEXONO}
The TEXONO collaboration took $0.338$~kg-days of data with a very-low threshold germanium detector~\cite{Lin:2007ka} in the range 
$E^\prime=0.1\sla 7.1\kevee$. The experiment was running between January 15 and February 6, 2007~\cite{texonopc}.
We obtained a data sample from the TEXONO collaboration from which we extracted the number of events in bins of size 
$100$~eVee\footnote{This data set can be obtained from the TEXONO
collaboration directly.}. 
The quenching factor is $Q=0.2$ and we again use only the bins up to $2\kevee$.

The overall efficiency is estimated as (cf.~Figure~3 in~\cite{Lin:2007ka})
\begin{align}
\xi(E^\prime)&=    0.983\cdot 0.915 \cdot \frac{2}{\pi}\,\arctan{\left(\frac{487\,{E^\prime}^4}{\mathrm{keVee}^4}\right)}  \;,
\label{eqn:TEXONOeff}
\end{align}
while the energy resolution can be approximated by
\begin{align}
\sigma(E')&=   \left( 0.055 + 0.009 \sqrt{\frac{E^\prime}{\mathrm{keVee}}} \right)\kevee \;.
\label{eqn:TEXONOER}
\end{align}

\section{Results}\label{sec:results}
\begin{figure}[t] 
  \centering
\hspace{-6mm}
  \begin{minipage}[b]{7.2 cm}
    \includegraphics[width=7.45cm]{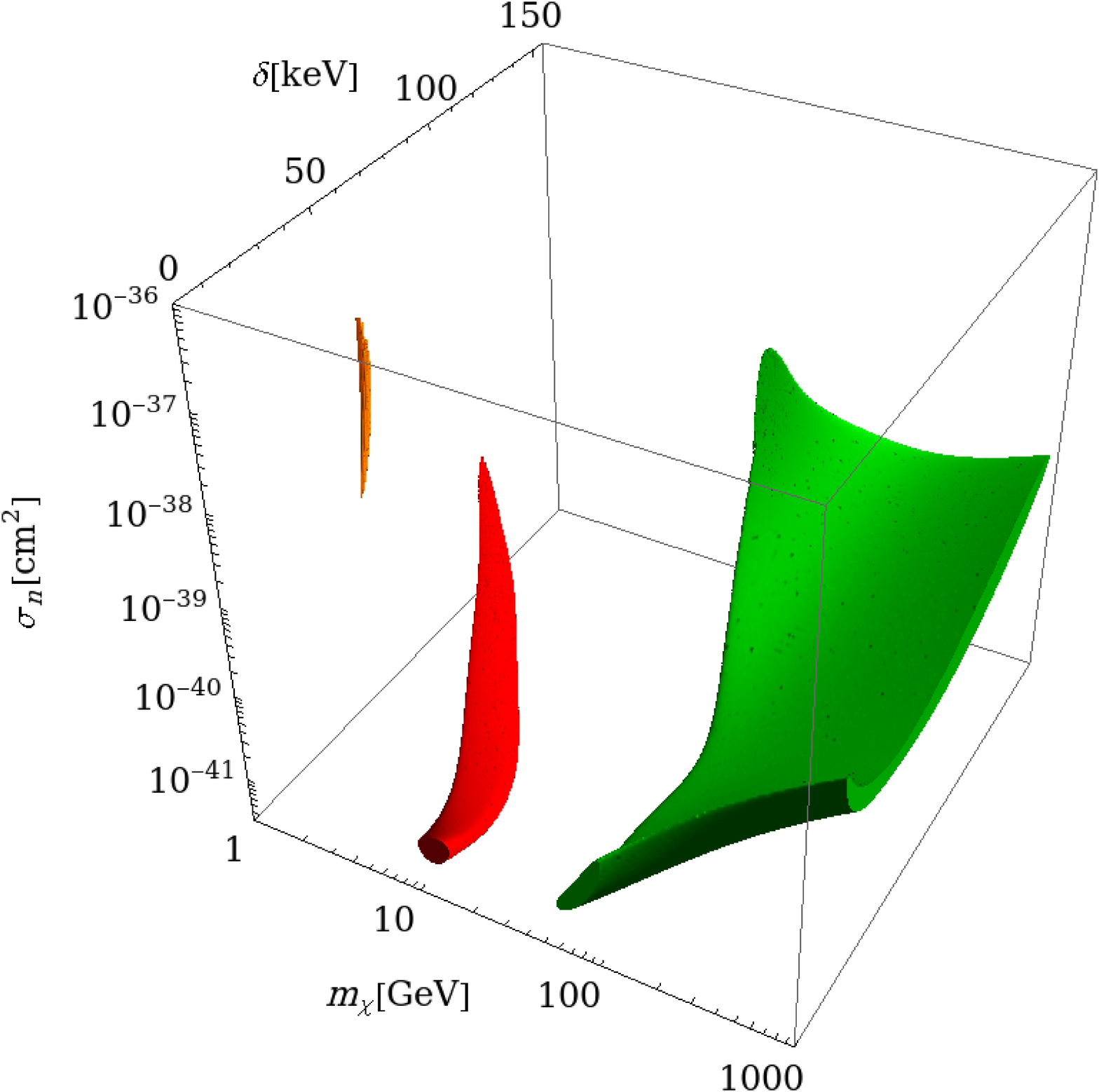}  
  \end{minipage}\hspace*{0.7cm}
  \begin{minipage}[b]{7.2 cm}
    \includegraphics[width=7.45cm]{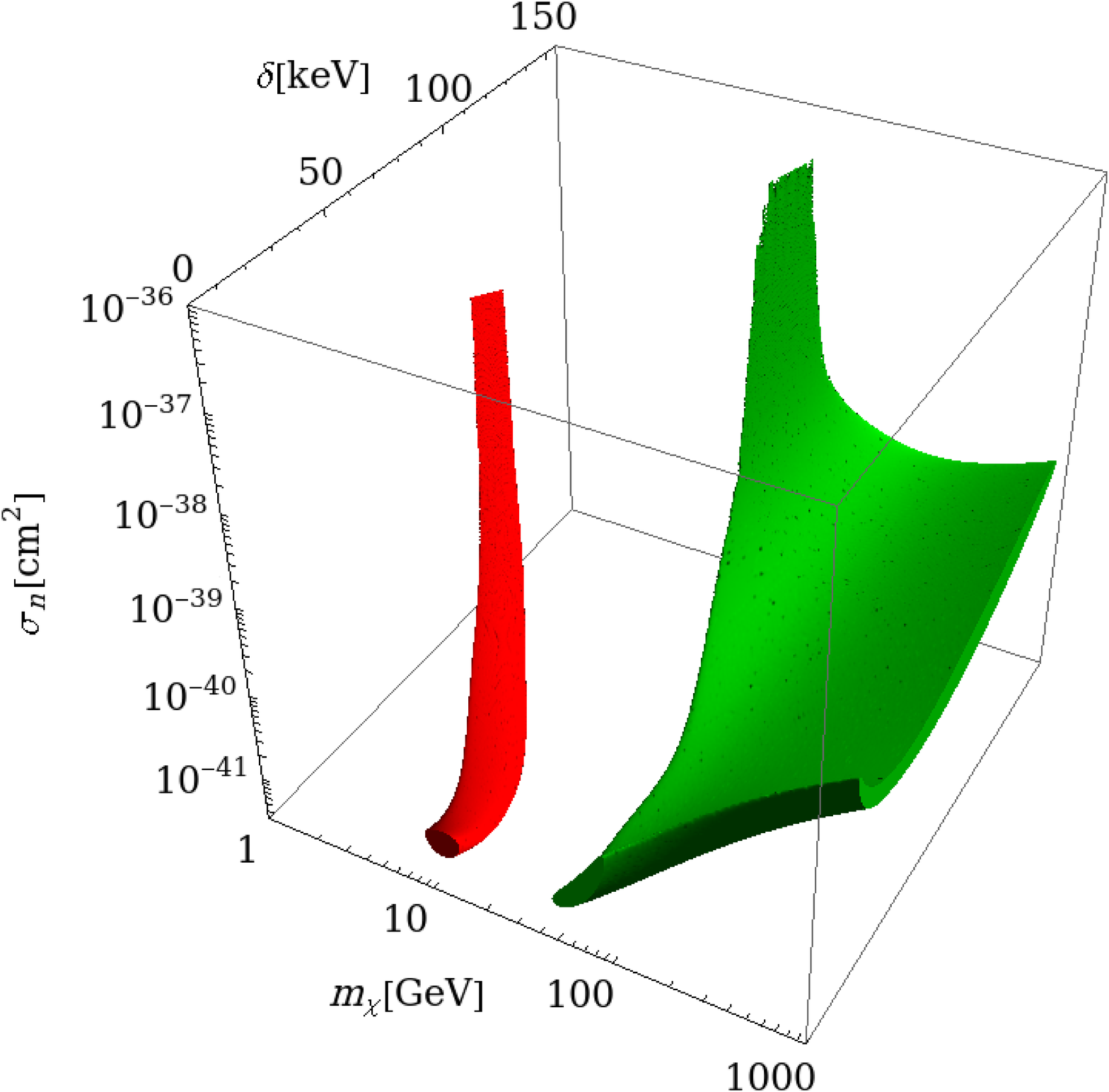}  
  \end{minipage}
  \caption{Parameter space of inelastic dark matter potentially explaining the DAMA annual modulation signal
           at 90\% CL without (left panel) and with (right panel) detector resolution of DAMA taken into account.
           For the color code see text.
}
  \label{fig:dama3d}
\end{figure}

In this section we describe the results of our analysis, i.e.~we show the regions in the parameter space of inelastic 
dark matter which are compatible with the DAMA signal and check if they are also consistent with the null results of the 
various other searches described in Section~\ref{sec:experiments}. 
The parameter space is spanned by three parameters: the WIMP mass $m_\chi$, 
the mass splitting $\delta$ and the WIMP neutron cross section $\sigma_n$.

We find three regions\footnote{If we enhance the confidence level, the distance between the red 
and the green region in Figure~\ref{fig:dama3d} decreases, at the 99\% level they are already slightly connected.} 
in parameter space which could potentially explain the DAMA signal. 
They are depicted in different colors in Figure~\ref{fig:dama3d}.
To illustrate the impact of the DAMA energy resolution, the left and right panel show the DAMA allowed 
regions without and with the resolution taken into account, respectively. While the large green and red regions are extended 
through the inclusion of the resolution, the small yellow region completely disappears. We will now discuss these three regions 
in turn and confront them with the limits from the various null searches. In this analysis we will
not consider additional constraints e.g.~from collider searches or indirect detection~\cite{Nussinov:2009ft,Menon:2009qj} 
as their applicability strongly depends on the considered model.

\subsection{The quenched region}
\begin{figure}[t] 
  \centering
  \begin{minipage}[b]{7.2 cm}
    \includegraphics[height=4.7cm]{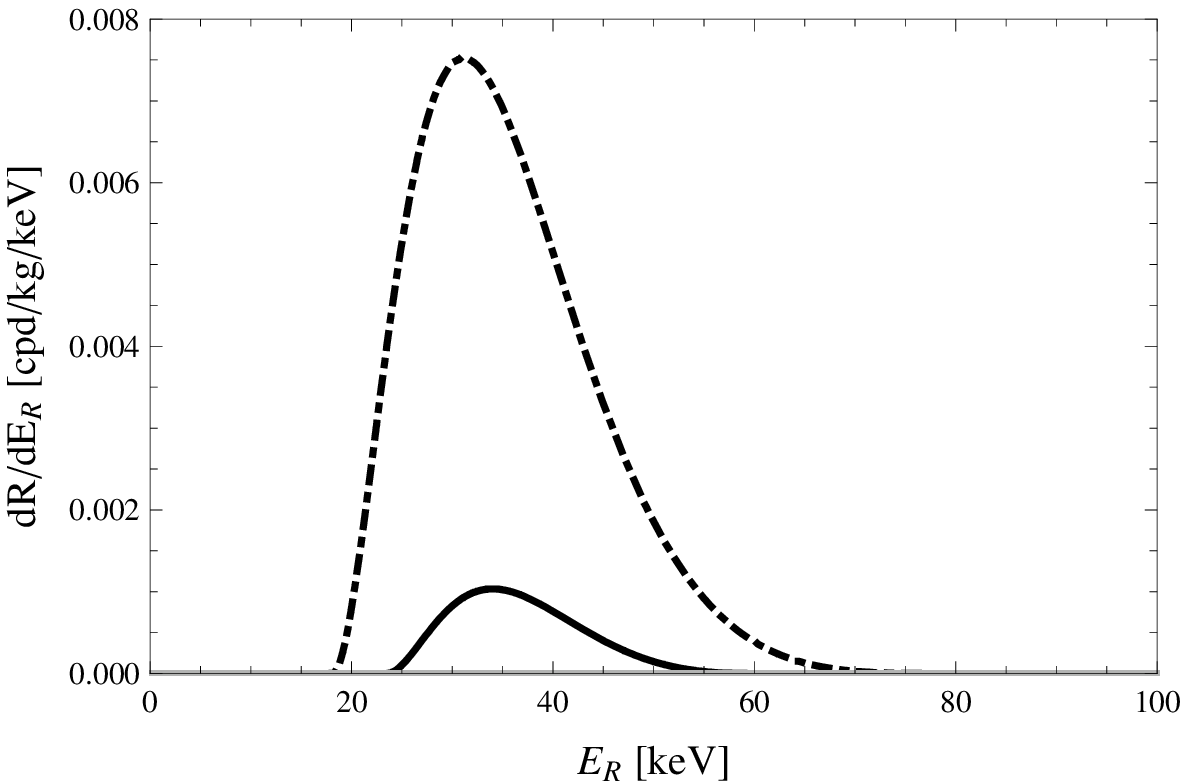}  
  \end{minipage}\hspace*{1cm}
  \begin{minipage}[b]{7.2 cm}
    \includegraphics[height=4.8cm]{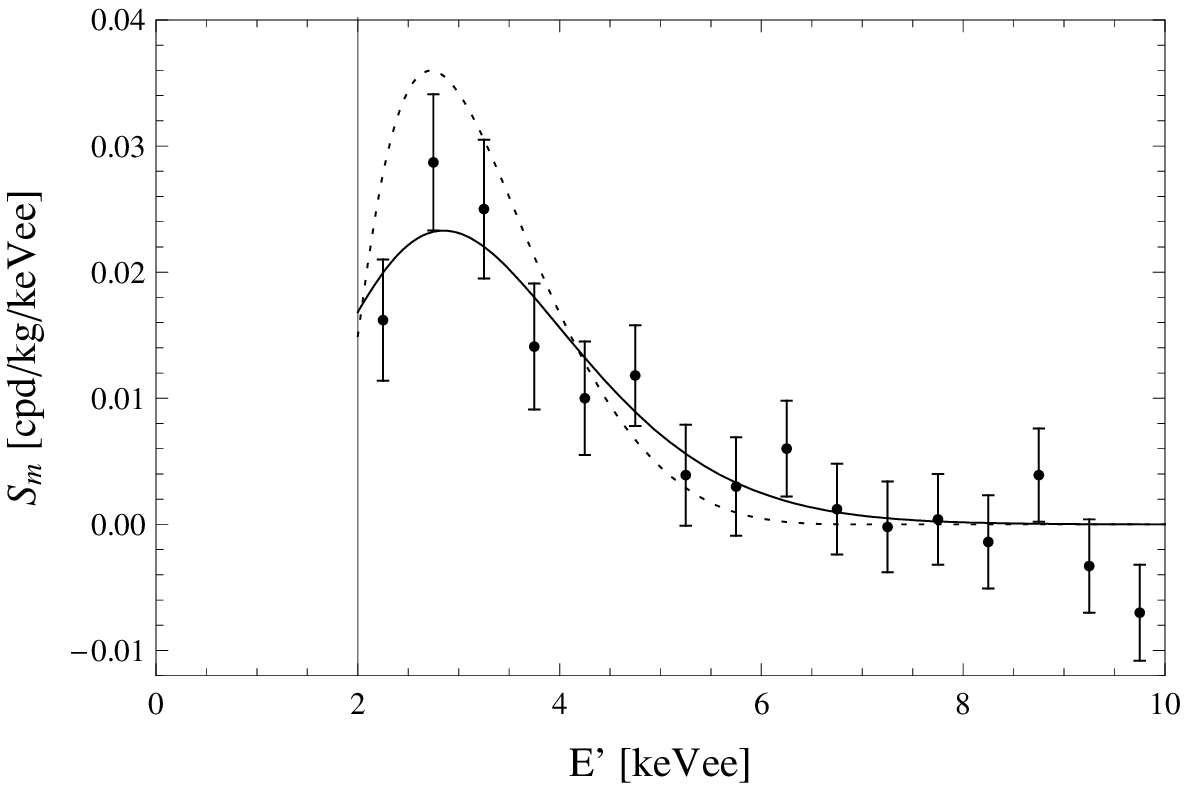}  
  \end{minipage}
  \caption{Recoil spectrum and  modulation amplitude at DAMA for the local best fit point
           $m_\chi=56.5\gev$, $\delta=115.4\kev$, $\sigma_n=2.5 \cdot 10^{-39}\cm^2$.
           Left Panel: Differential event rate $\mathrm{d}R/\mathrm{d}E_R$ in a NaI target as a function of the recoil energy 
           $E_R$ for the parameters above. Only iodine events are visible as scattering off of sodium is kinematically not 
           accessible. The dash-dotted curve refers to June 2, the solid curve to December 2.
           Right Panel: Predicted modulation amplitude at DAMA for measured energies $E^\prime$ up to $10\kevee$ 
           with (solid) and without (dashed) resolution taken into account. The modulation amplitude measured by DAMA is 
           also included.
}
  \label{fig:dama100gev}
\end{figure}

The quenched iodine region corresponds to the green region in Figure~\ref{fig:dama3d} with 
$m_\chi \gtrsim 43\gev$ and $0 \lesssim \delta \lesssim 145\kev$ when the detector resolution is taken into account. 
While the plot ends at $m_\chi = 1\tev$, the region extends to extremely high masses. 
Allowed cross sections range from $\sigma_n \simeq 6\cdot 10^{-42}\cm^2$ to $10^{-36}\cm^2$, 
a very thin strip extends above  $10^{-36}\cm^2$.

Only quenched events can contribute to the signal. This can be seen in the 
left panel of Figure~\ref{fig:dama100gev} where the recoil spectrum in sodium iodide for one of the two local best 
fit points\footnote{The other local best fit point lies at $\delta \sim 0$, it has a slightly lower value of $\chi^2$.} 
($m_\chi=56.5\gev$, $\delta=115.4\kev$ and $\sigma_n=2.5 \cdot 10^{-39}\cm^2$ with $\chi^2=12.1$) is shown. 
The lower threshold of the iodine spectra lies at recoil energies $E_R\sim 20\kev$, i.e.~there are no channeled 
events in the relevant energy range below $\sim 10\kevee$.

In the right panel of Figure~\ref{fig:dama100gev} the predicted modulation amplitude (with and without taking 
into account the detector resolution) is shown for the same parameters together with the measured 
modulation amplitude in the bins up to $10\kevee$. Note that the finite energy resolution leads to a slight smearing of the 
spectrum. Especially parameter sets which predict too high spectra at zero resolution may become compatible with DAMA through 
this smearing. This is the reason why the DAMA allowed green region in Figure~\ref{fig:dama3d} extends to larger 
$\sigma_n$ and smaller $m_\chi$ if the resolution is 
taken into account.

\begin{figure}[t] 
  \centering
  \begin{minipage}[t]{7.2 cm}
    \includegraphics[width=7cm]{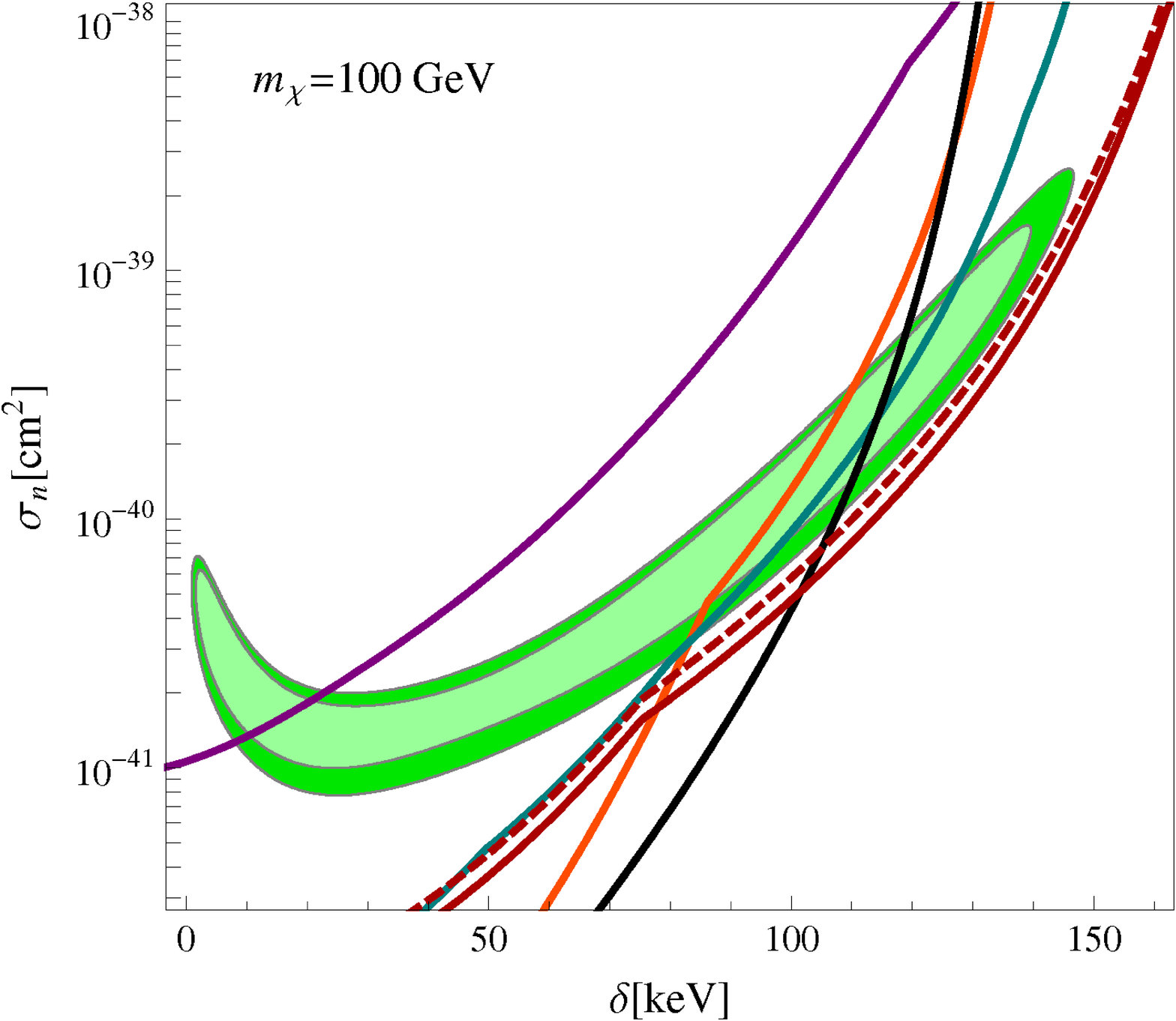}  
  \end{minipage}
\hspace{10mm}
  \begin{minipage}[t]{7.2 cm}
    \includegraphics[width=7cm]{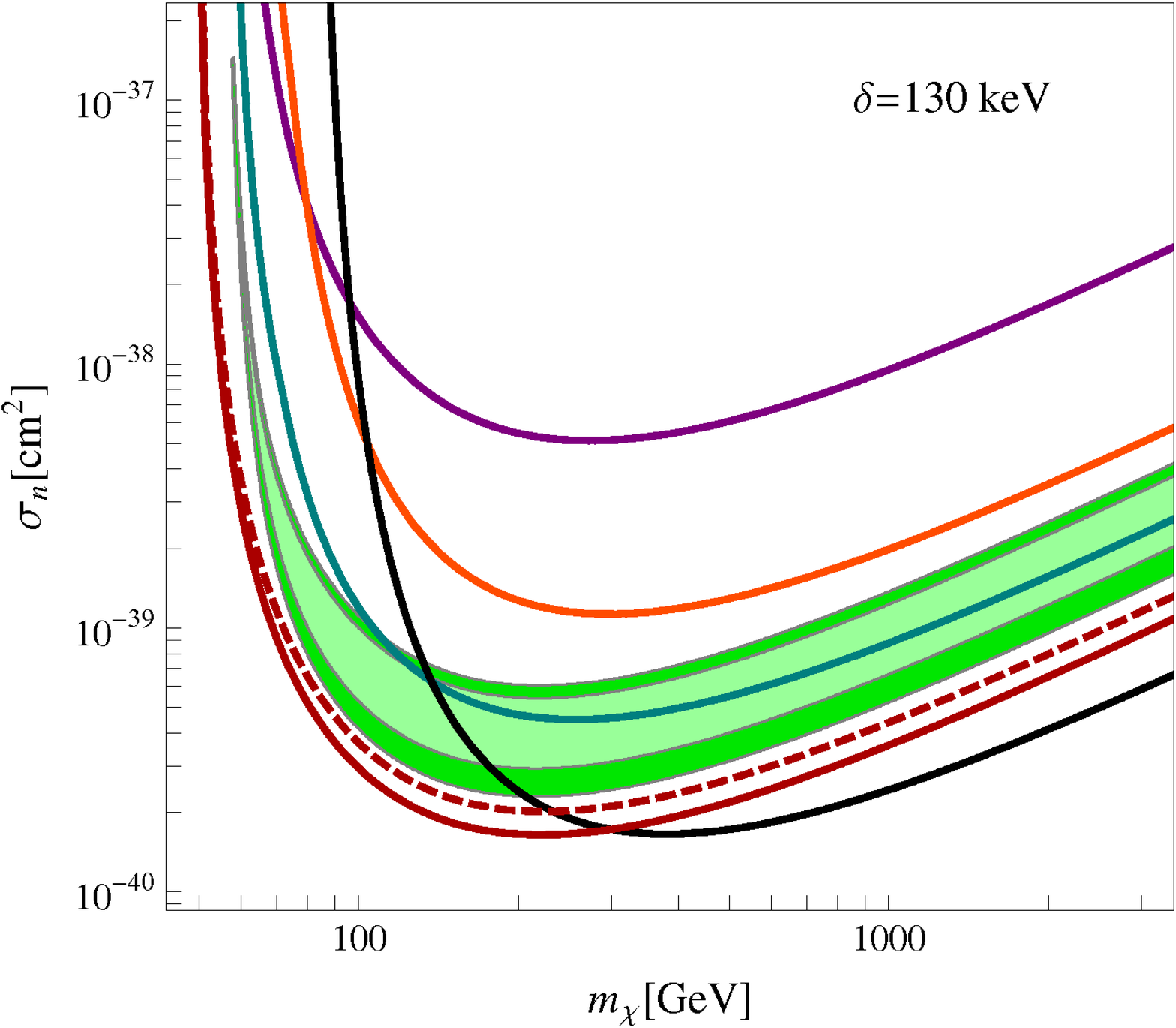}  
  \end{minipage}\\ 
\hspace{5mm}
    \includegraphics[width=4cm]{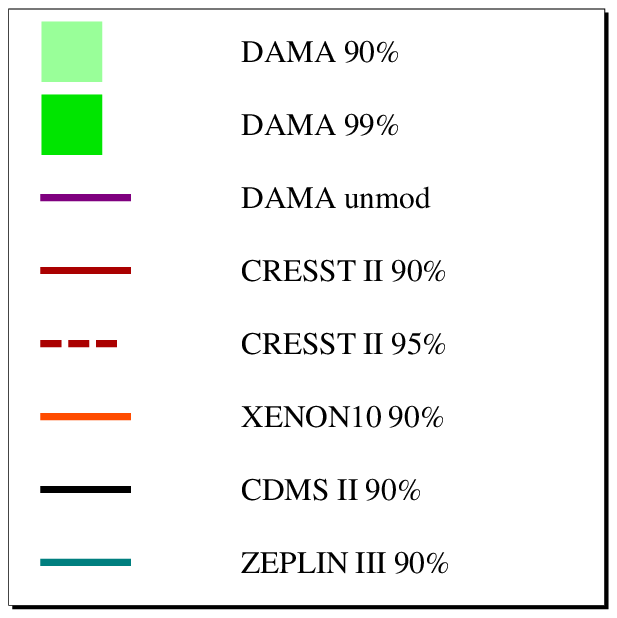}  
  \caption{DAMA allowed regions in the parameter space of inelastic dark matter together with the relevant constraints 
           from other direct detection experiments for fixed $m_\chi=100\gev$ (left panel) and fixed $\delta=130\kev$ (right panel).
           The region above the lines is excluded.
}
  \label{fig:quenched}
\end{figure}

In Figure~\ref{fig:quenched} the limits from the DAMA absolute rate and the other direct detection experiments on 
the quenched iodine region are shown. 
We plot only constraints from CRESST~II, ZEPLIN~III, CDMS~II and XENON10, as constraints from other experiments are significantly weaker 
in this regime. We fix $m_\chi=100\gev$ in the left panel and $\delta=130 \kev$ in the right panel and scan over the two remaining 
parameters. For DAMA we show the 90\% and 99\% allowed regions, while all exclusion curves are given at 90\% CL. 
Only for CRESST~II, which gives the most important constraint, we also include the 95\% exclusion curve.
In fact 
CRESST~II excludes the whole 99\% DAMA allowed region at 95\% CL. 
We checked that this statement holds not only for the parameters chosen above, but in the complete quenched iodine region 
over the whole range of $m_\chi$ and $\delta$. 

We then tested the robustness of this statement by varying the iodine quenching factor $Q_\mathrm{I}$ and the local 
galactic escape velocity 
$v_\text{esc}$ within their experimental uncertainties. While high $v_\text{esc}=600$~km/s and low $v_\text{esc}=500$~km/s did not 
weaken the CRESST~II bound, the latter becomes slightly weaker for the quenching factor at its minimal or maximal values
of  $Q_\mathrm{I}=0.08$ and $Q_\mathrm{I}=0.1$~\cite{iquench} respectively. Nevertheless, 
even if we tune $Q_\mathrm{I}$ in the most favorable way, the DAMA 99\% confidence region is still excluded at $90\%$ 
CL by CRESST~II in the whole quenched iodine region.

One should, however, be aware that we derived our limits for the Standard Halo Model. If the velocity distribution of WIMPs shows 
substantial deviations from this model, there may exist the possibility to reduce the level of incompatibility between DAMA and 
CRESST~II. Nevertheless tungsten is kinematically favored over iodine in the quenched iodine regime, i.e.~there are no parts in 
the velocity distribution of WIMPs which are accessible to DAMA but not to CRESST~II. Consequently there is no simple way to 
resolve the tension between the two.

\subsection{The channeled regions}

\subsubsection{Sodium}

\begin{figure}[t] 
  \begin{minipage}[t]{7.2 cm}
    \includegraphics[height=4.8cm]{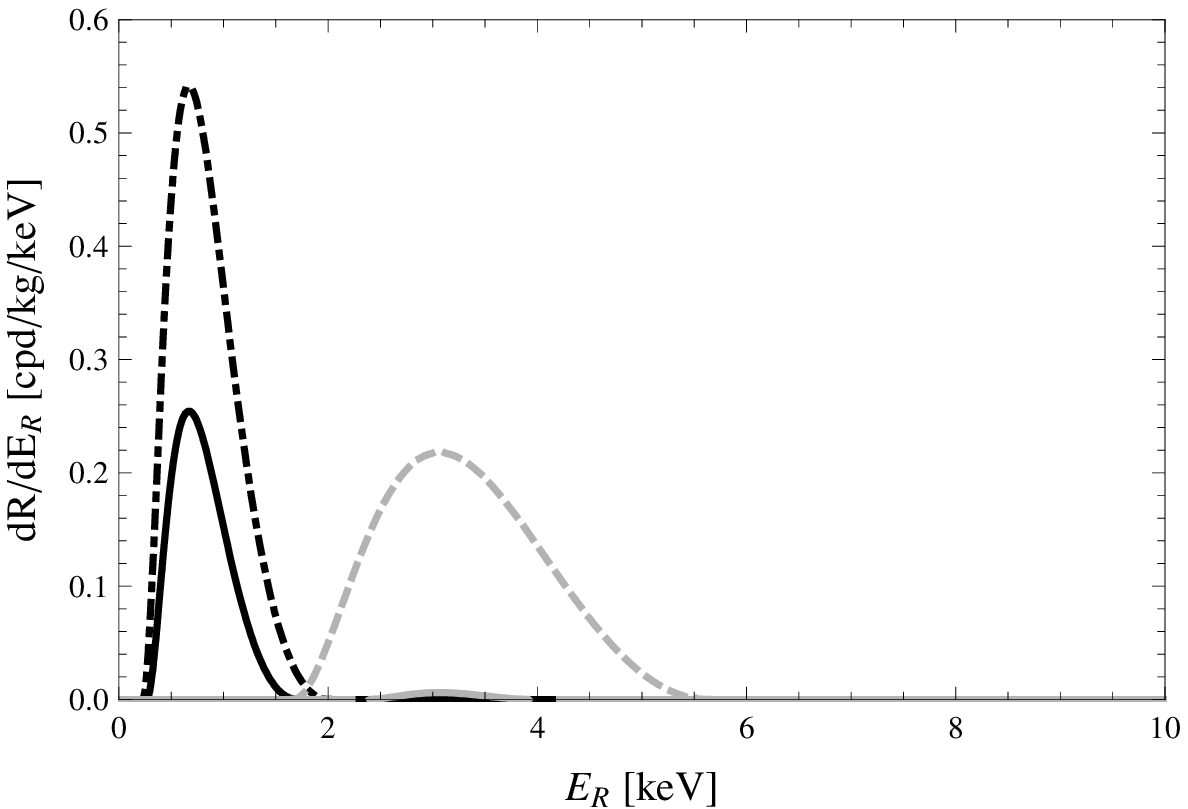}  
  \end{minipage}\hspace*{1cm}
  \begin{minipage}[t]{7.2 cm}
    \includegraphics[height=4.8cm]{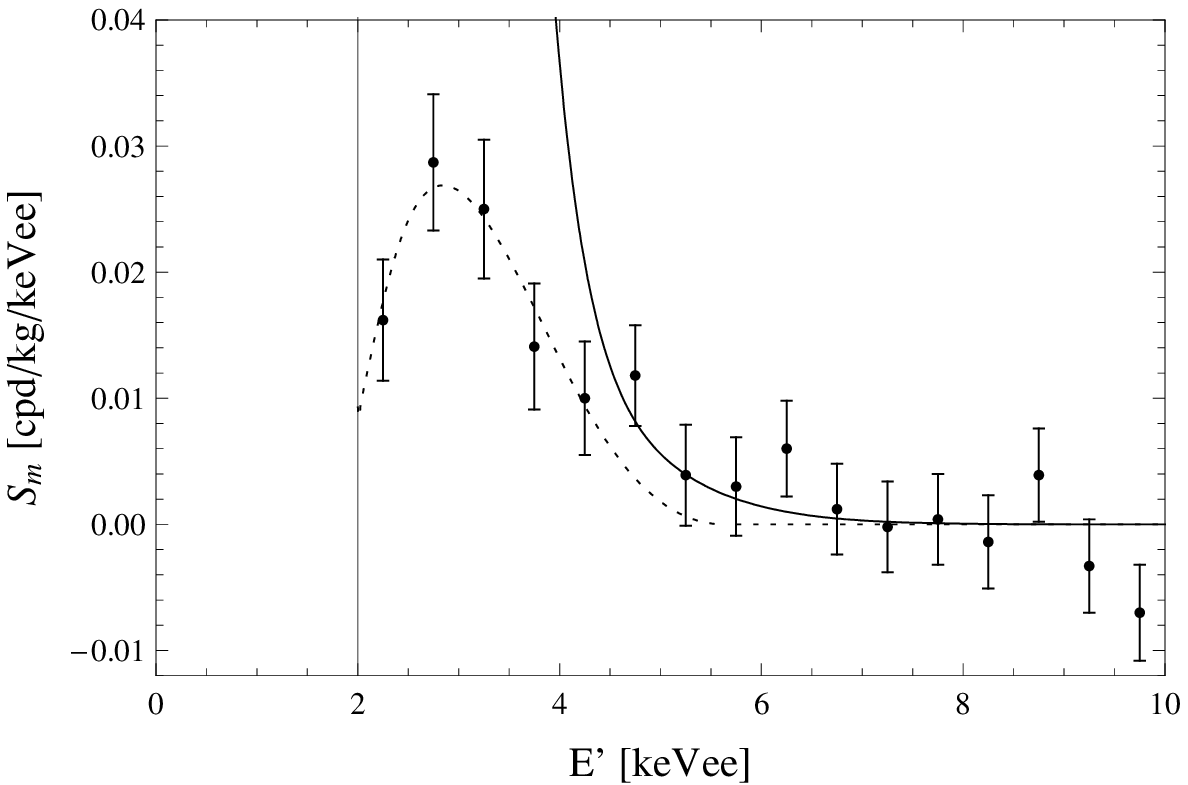} 
  \end{minipage}   
  \caption{Same as Figure~\ref{fig:dama100gev} but for 
           $m_\chi=5.82 \gev$, $\delta= 14.34\kev$, $\sigma_n= 3.9 \cdot 10^{-37}\cm^2$ (best fit point in the channeled sodium 
           region at zero resolution). The gray curves in the left panel refer to sodium, the black curves again to iodine. 
           Note that the iodine spectrum has been rescaled by a factor 1/1000.
}
  \label{fig:dama5gev}
\end{figure}

The channeled sodium region which corresponds to the yellow region in the left panel of Figure~\ref{fig:dama3d} lies at 
$5.3 \gev \lesssim m_\chi \lesssim6.1 \gev$, $10\kev \lesssim \delta \lesssim 16\kev$ and $2\cdot 10^{-38}\cm^2 \lesssim 
\sigma_n \lesssim 2 \cdot 10^{-36}\cm^2$ for the detector resolution set to zero. 
It is the only region where the DAMA modulation amplitude can potentially be 
explained by scatterings off of sodium. The reason why this region disappears when we take into 
account the resolution can easily be inferred from the left panel of Figure~\ref{fig:dama5gev}. There we show the 
sodium and the iodine recoil spectra for the best fit point at zero resolution ($m_\chi=5.82 \gev$, $\delta= 14.34\kev$ and 
$\sigma_n= 3.9 \cdot 10^{-37}\cm^2$ with $\chi^2=15.5$). 
One can see that there is an extremely large number of iodine events below the DAMA 
threshold which in case of a nonzero resolution are partly shifted to higher energies. 
In the right panel of the same figure we see 
that this totally spoils the fit to the DAMA modulation amplitude.

Nevertheless we hesitate to completely exclude the sodium channeled region. In some part of it, the iodine peak lies at energies 
$E^\prime \lesssim 1.5\kevee$. In such cases the compatibility with DAMA could remain if $\sigma(1.5\kevee) \lesssim 0.2\kevee$ 
which is about $1/3$ of the expected resolution of DAMA/LIBRA at that energy. One should have in mind that the resolutions 
of DAMA/NaI and DAMA/LIBRA were only extrapolated to such low energies which implies some uncertainties. In addition, 
the resolution may show some deviations from Gaussianity far away from the central value. Finally there is also the possibility that 
the iodine peak is somehow affected by DAMA's noise rejection procedure near threshold. 

On the other hand, the total rate at DAMA/LIBRA (Figure~1 in~\cite{Bernabei:2008yi}) shows no sharp 
rise above $E^\prime=0.75\kevee$. And while one should be very careful about the interpretation of bins below the threshold, 
there is at least no indication for the iodine peak down to this energy.
Although the existence of the sodium channeled region is therefore doubtable, we nevertheless present here the limits 
on this region from other direct detection experiments. For this purpose we set the energy resolution of DAMA to zero. 
Note that we do this just for illustration, the definite position and size of the DAMA allowed region, if it exists,
highly depends on the behavior of the DAMA detector below its threshold.

\begin{figure}[t] 
  \centering
  \begin{minipage}[t]{7.2 cm}
    \includegraphics[width=7.05cm]{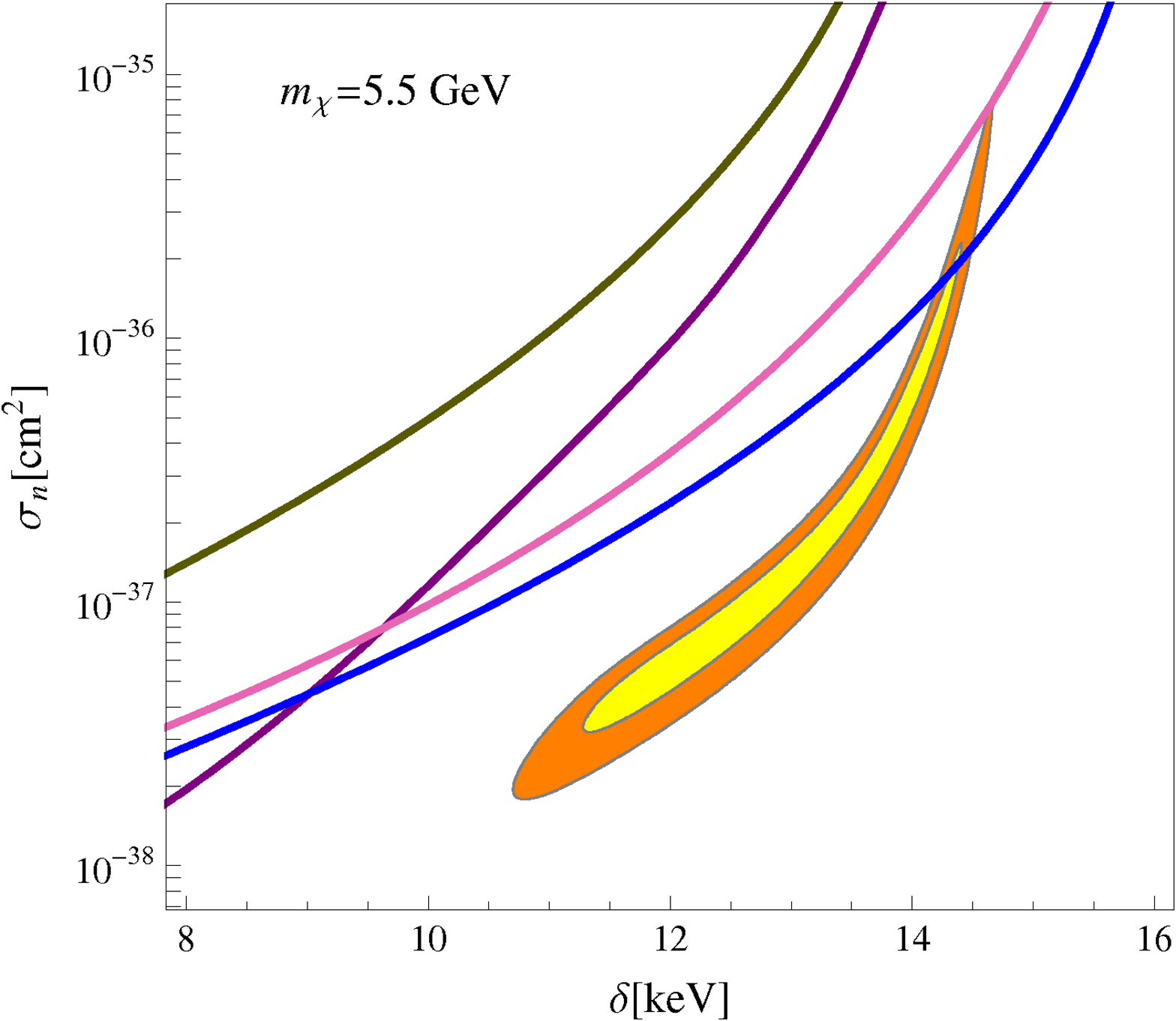}  
  \end{minipage}\hspace*{1cm}
  \begin{minipage}[t]{7.2 cm}
    \includegraphics[width=7.0cm]{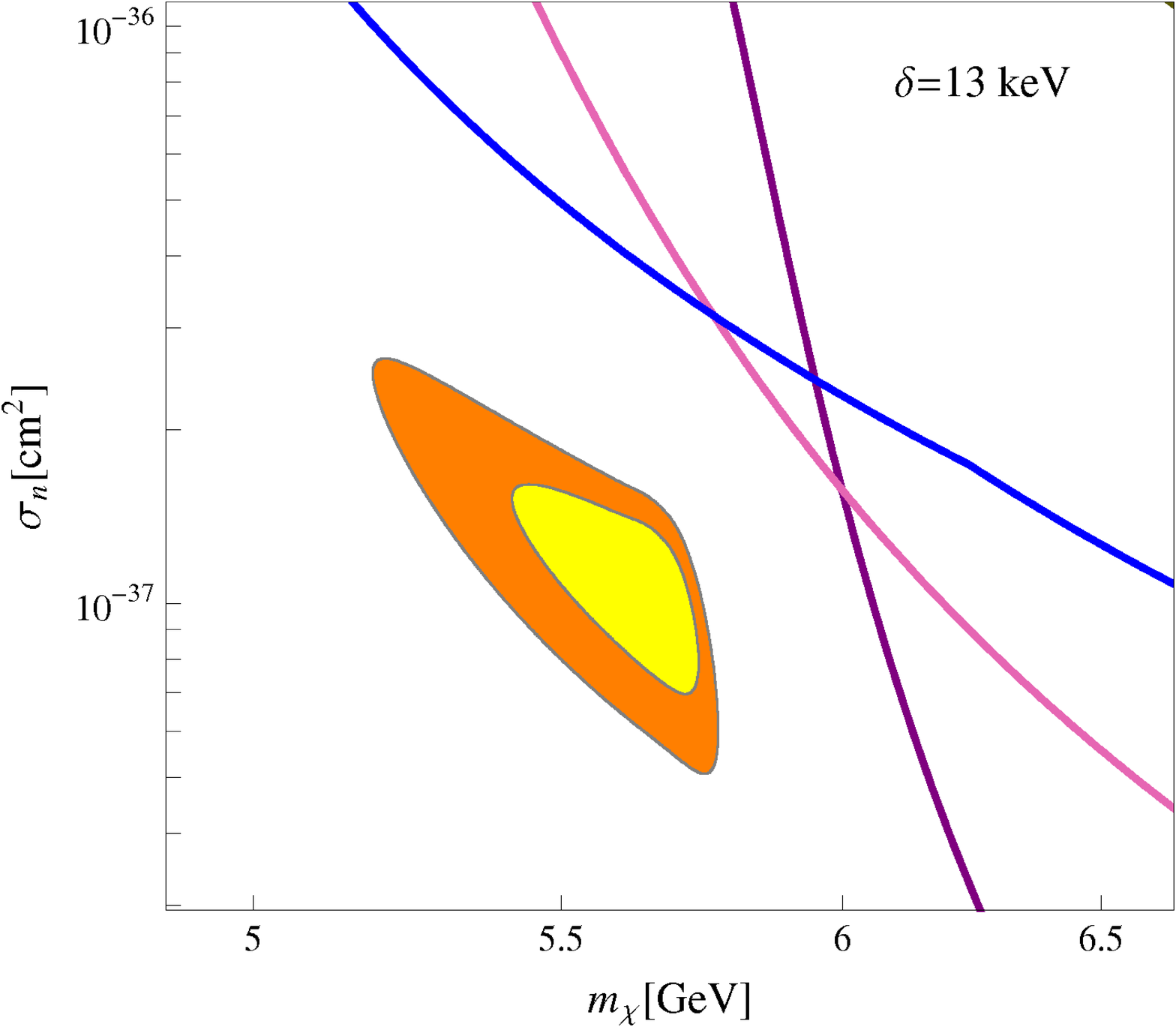}  
  \end{minipage}  \\ 
\hspace{8mm}
    \includegraphics[width=4cm]{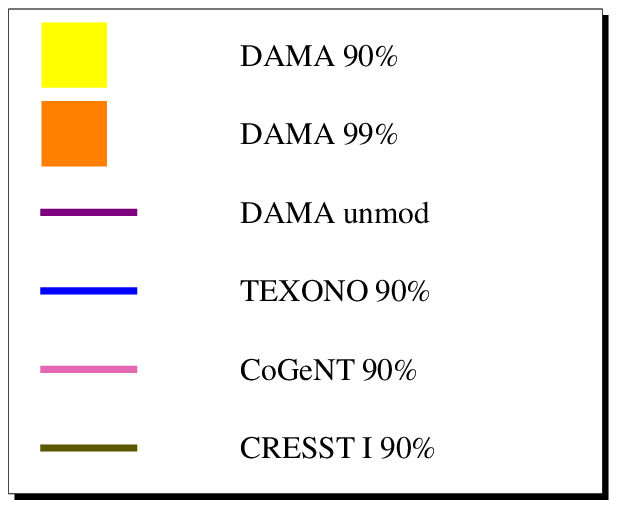}  
\caption{Same as Figure~\ref{fig:quenched}, but for $m_\chi=5.5\gev$ (left panel) and $\delta=13\kev$ (right panel) and the detector resolution of DAMA set to zero (see text).}
  \label{fig:channeledNa}
\end{figure}

As the sodium region lies at very low $m_\chi$, important constraints only arise from the very low-threshold experiments 
CoGeNT, TEXONO and CRESST~I. In Figure~\ref{fig:channeledNa} we show the DAMA allowed regions (90\% and 99\%) at zero 
resolution together with the 90\% exclusion curves from these experiments for fixed $m_\chi=5.5\gev$ (left panel) or fixed 
$\delta=13 \kev$ (right panel). 
As can be seen only TEXONO has begun to explore a small part of the DAMA allowed region, most of it still remains unchallenged.

\subsubsection{Iodine}
\begin{figure}[t] 
  \begin{minipage}[b]{7.2 cm}
    \includegraphics[height=4.7cm]{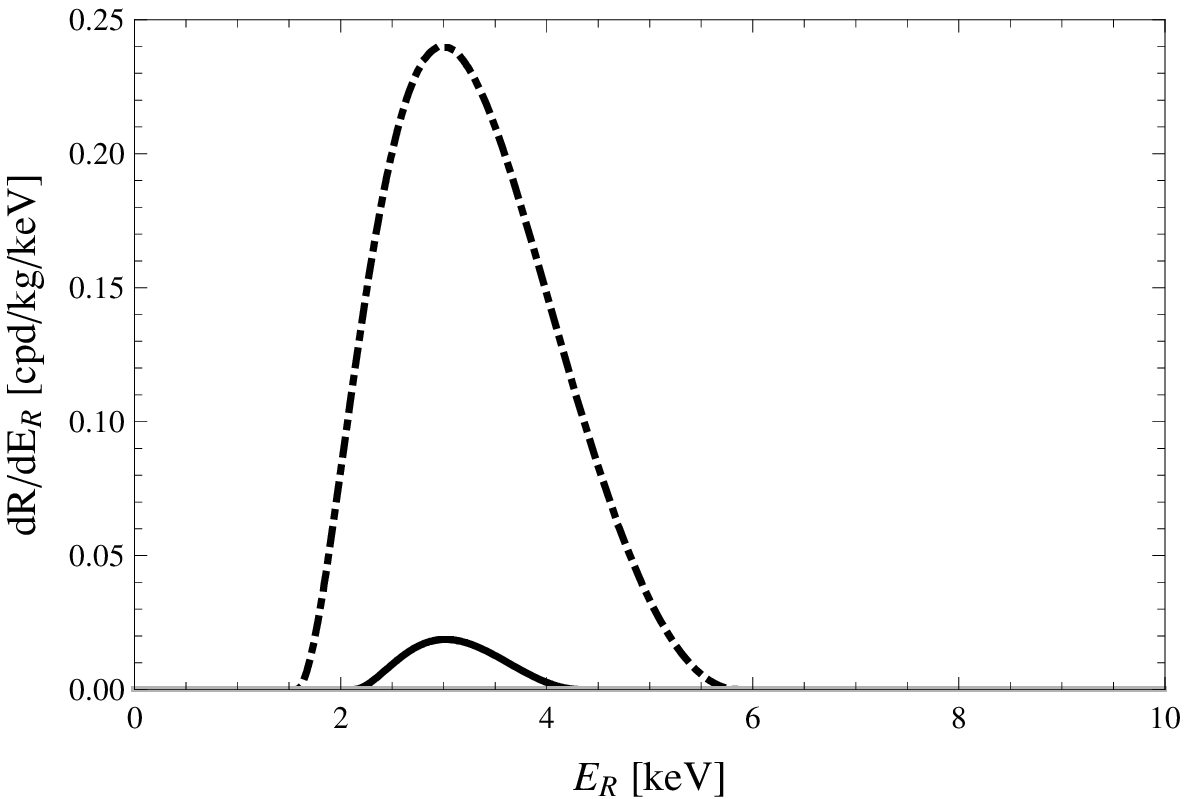}  
  \end{minipage}\hspace*{1cm}
  \begin{minipage}[b]{7.2 cm}
    \includegraphics[height=4.8cm]{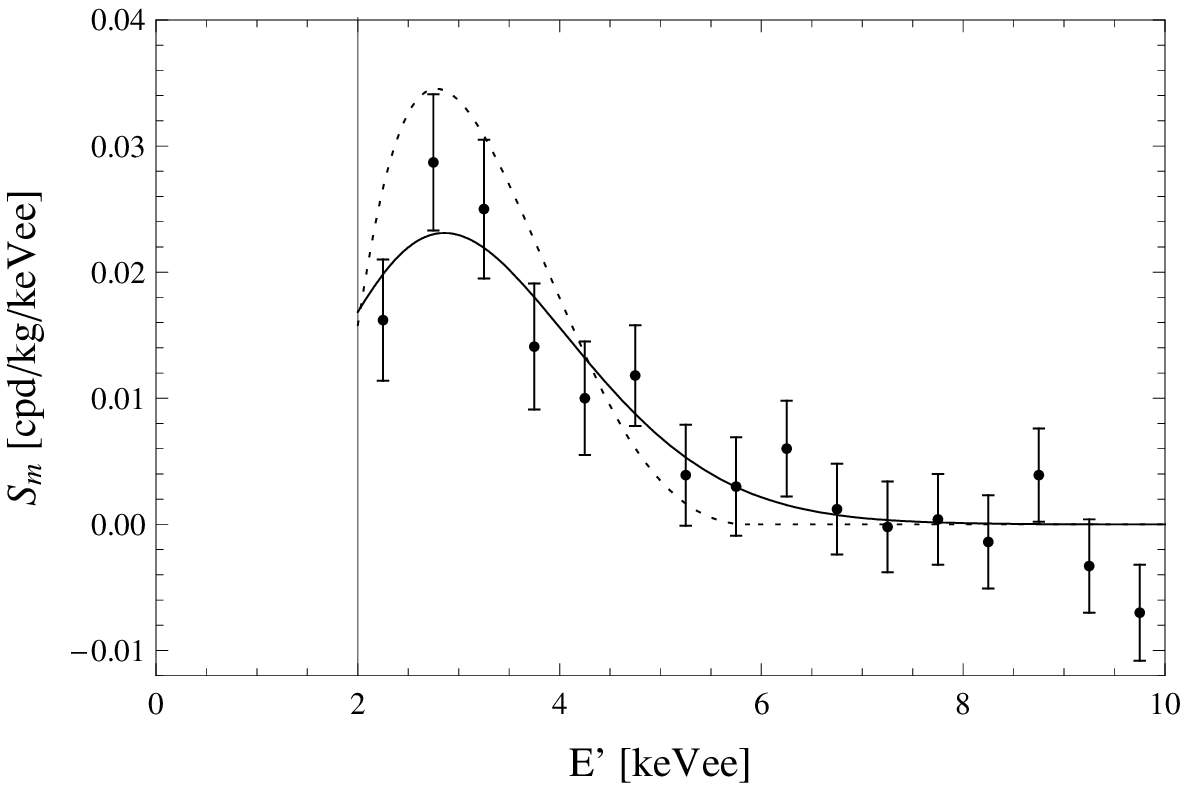}  
  \end{minipage}
  \caption{Same as Figure~\ref{fig:dama100gev} but for the best fit point in the channeled iodine region 
           ($m_\chi=11.9\gev$, $\delta=33.3\kev$, $\sigma_n=4.7 \cdot 10^{-39}\cm^2$). All spectra refer to iodine as 
           scattering off of sodium is kinematically not accessible. 
}
  \label{fig:dama12gev}
\end{figure}

The channeled iodine region corresponds to the red region in Figure~\ref{fig:dama3d} with $9 \gev \lesssim m_\chi \lesssim20 \gev$ 
and $0 \lesssim \delta \lesssim 38\kev$ when the detector resolution is taken into account. 
Allowed cross sections range from $\sigma_n \simeq 9\cdot 10^{-42}\cm^2$ to $10^{-36}\cm^2$, a very thin strip extends 
above  $10^{-36}\cm^2$. 
Only channeled iodine events give a contribution to the signal. This can be seen in the left panel of Figure~\ref{fig:dama12gev} 
where we plot the recoil spectrum in sodium iodide for the best fit point 
($m_\chi=11.9\gev$, $\delta=33.3\kev$ and $\sigma_n=4.7 \cdot 10^{-39}\cm^2$ with $\chi^2=12.3$). 
As the iodine spectrum is practically zero above $E_R\sim 6\kev$, all quenched events lie below the DAMA threshold and 
cannot contribute to the signal.

The right panel shows the predicted modulation amplitude at the best fit point and the measured modulation amplitude in the bins up to $10 \kevee$. 
As in the quenched iodine region the inclusion of the energy resolution causes a slight smearing of the spectrum which leads 
to an extension of the DAMA allowed region.

\begin{figure}[t] 
  \centering
  \begin{minipage}[t]{7.2 cm}
    \includegraphics[width=7.0cm]{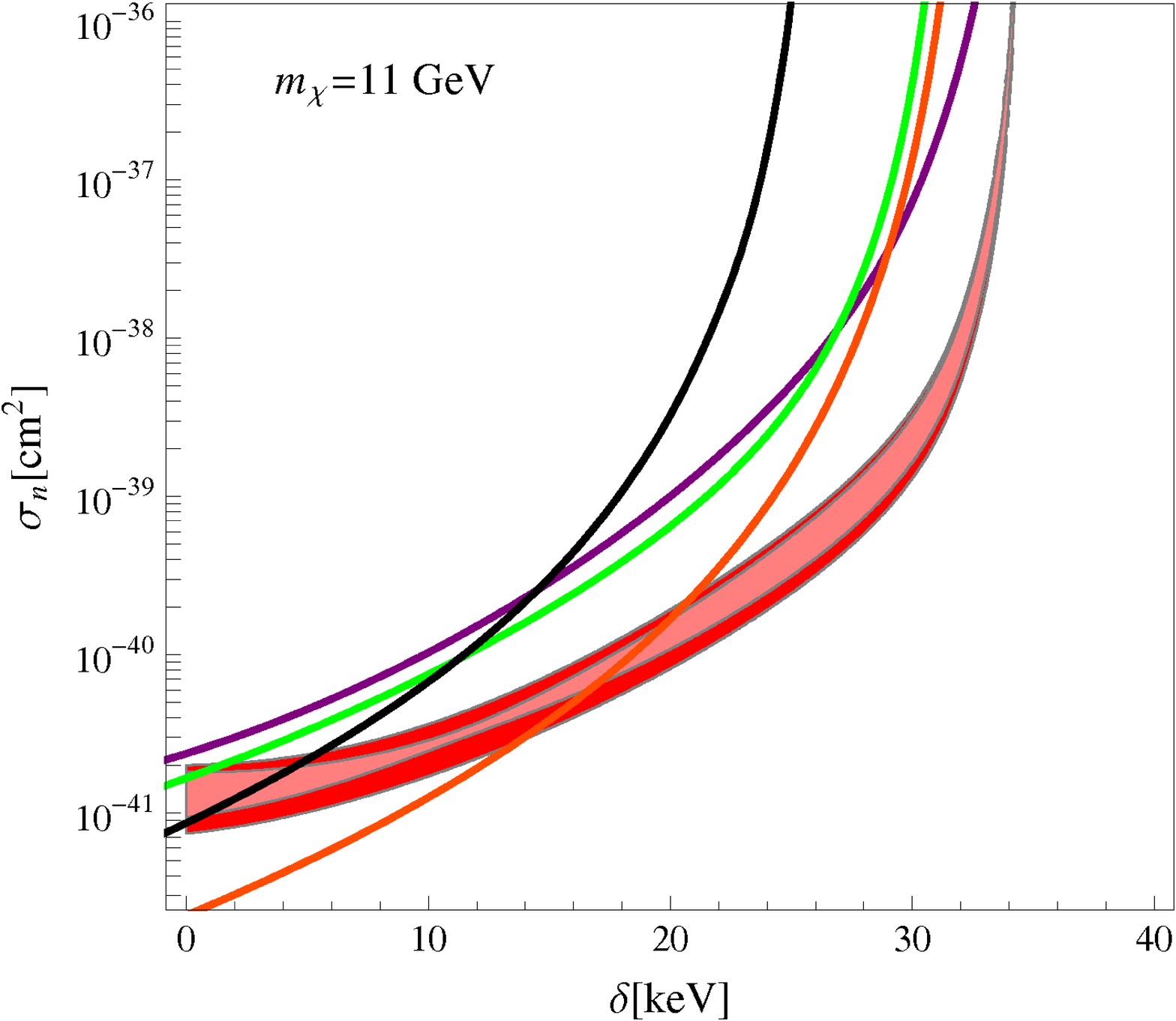}  
  \end{minipage}\hspace*{1cm}
  \begin{minipage}[t]{7.2 cm}
    \includegraphics[width=7cm]{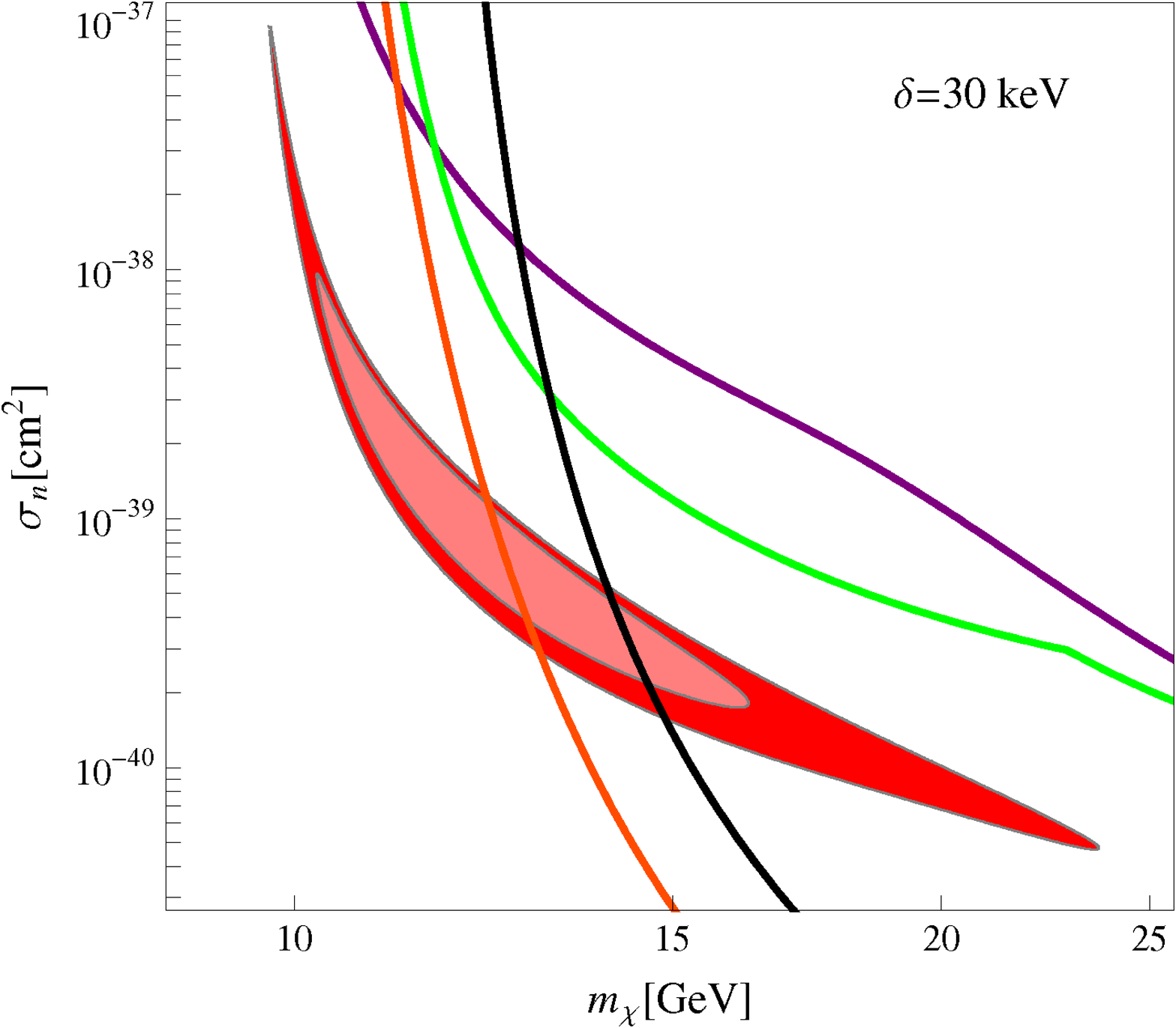}  
  \end{minipage}\\ 
\hspace{9mm}
    \includegraphics[width=4cm]{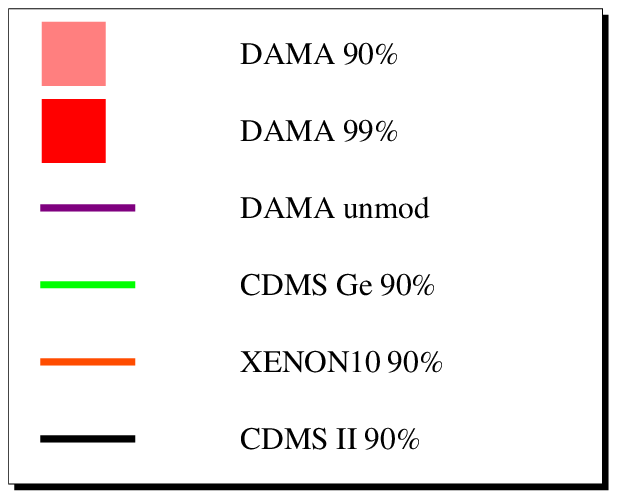}  

  \caption{Same as Figure~\ref{fig:quenched}, but for $m_\chi=11\gev$ (left panel) and $\delta=30\kev$ (right panel).
}
  \label{fig:delta2}
\end{figure}

The most important constraints on the channeled iodine region arise from XENON10 and CDMS~II,
as can be seen in Figure~\ref{fig:delta2}. 
In particular, we find that the part of the channeled iodine region where $m_\chi\gtrsim15\gev$ is excluded by XENON10 at 99\% CL. 
However, below this mass there still exists parameter space in agreement with DAMA and completely consistent with all other experiments.

\section{Comparison with other studies}\label{sec:comparisonwithother}
  
Our results show important deviations from earlier studies~\cite{Chang:2008gd,Cui:2009xq,MarchRussell:2008dy} which we 
now want to address in some detail. We focus on heavy inelastic dark matter (green region in Figure~\ref{fig:dama3d}) 
as the low mass regions have not been studied systematically.

In contrast to~\cite{Chang:2008gd,Cui:2009xq,MarchRussell:2008dy} we took into account the energy resolution of 
DAMA/LIBRA and DAMA/NaI which leads to an extension of the DAMA allowed region especially to larger cross sections and smaller masses 
(c.f.~Figure~\ref{fig:dama3d}). 
We also took into account information from the higher energy bins.

As for the other experiments, the most important difference concerns the constraint from CRESST~II which is more severe in 
our case than in~\cite{Chang:2008gd,Cui:2009xq,MarchRussell:2008dy}. In~\cite{Cui:2009xq} simple Poisson statistics were used
to derive the CRESST~II limit which allows a comparison between the absolute number of predicted and measured events only, 
but ignores information from their spectral shape. Interval based methods like the pmax-method employed 
in~\cite{Chang:2008gd,MarchRussell:2008dy} or the maximum gap method used by us seem more appropriate and lead to stronger limits. 
One would therefore expect that~\cite{Chang:2008gd,MarchRussell:2008dy} obtain comparably severe constraints. 
However, these studies combined the data set from the commissioning run of CRESST~II with the older test run.
In our analysis we refrain from including this run as explained in detail in Section~\ref{sec:cresst2}.  
But even if this run is included there is still some tension between CRESST~II and DAMA unless one particular event at $\sim 22\kev$  
measured by the Daisy module is included~\cite{MarchRussell:2008dy}. This event, however, enters the 
WIMP search region only if the quenching factor\footnote{Note that in cryogenic detectors the quenching factor 
has no impact on the energy calibration but only on the discrimination between background and candidate events.} of tungsten deviates 
from its experimental value (see discussion in~\cite{Angloher:2004tr}). Newer measurements of the quenching factor~\cite{Ninkovic:2006xy} 
seem to exclude this possibility, i.e.~the mentioned event should not be treated as a candidate event.

\section{Conclusions}\label{sec:conclusions}

In this study we examined the compatibility of the DAMA annual modulation signal with other direct searches 
for the case of inelastic dark matter.
We searched for regions consistent with the spectral shape of the DAMA modulation amplitude by scanning over the 
three-dimensional parameter space of inelastic dark matter which is spanned by the WIMP mass $m_\chi$, the mass splitting 
$\delta$ and the WIMP neutron cross section $\sigma_n$. 
We found two regions in parameter space which correctly reproduce the DAMA signal, one of them at higher $m_\chi$ 
corresponding to quenched and one at small $m_\chi$ corresponding to channeled scattering events off of iodine. 
A third region where the DAMA signal can be explained by channeled scatterings off of sodium opens up if 
we decrease the energy resolution of DAMA. Its existence depends strongly on the behavior of the DAMA detector below threshold.

All DAMA allowed regions were studied in detail and limits from the relevant other direct detection experiments were 
applied. 
In contrast to previous studies we found that for the Standard Halo Model inelastic dark matter with $m_\chi\gtrsim 15\gev$
is ruled out at 95\% CL where the strongest constraints are set by XENON10 (at smaller $m_\chi$) 
and CRESST II (at larger $m_\chi$). This statement remains fairly robust with respect to 
experimental and astrophysical uncertainties: when varying the galactic escape velocity and the iodine quenching factor 
within their experimental uncertainties, the DAMA 99\% confidence region remains excluded at $90\%$ CL. 
Dark matter velocity profiles which differ substantially from the Standard Halo Model 
may reduce the discrepancy between CRESST~II and DAMA in the high mass region slightly.

At lower masses, where the DAMA signal can be explained through channeled events at iodine and sodium respectively, 
we find that parts of these regions are consistent with all direct detection experiments.
In conclusion we have shown that heavy inelastic dark matter is disfavored by the combination of DAMA spectral information 
and the exclusion limits from other experiments. Light inelastic dark matter on the other hand constitutes
a viable solution to the DAMA puzzle.

\subsection*{Acknowledgments}
We would like to thank Michael Ratz and Walter Potzel for useful discussions as well as Henry Wong, Juan Collar, Michael Dragowsky 
and Franz Pr\"obst for correspondence. This research was supported by the DFG cluster of excellence "Origin and Structure of the 
Universe", and the \mbox{SFB-Transregio} 27 "Neutrinos and Beyond" by Deutsche
Forschungsgemeinschaft (DFG).

\bibliography{sneubib}

\providecommand{\bysame}{\leavevmode\hbox to3em{\hrulefill}\thinspace}
\frenchspacing
\newcommand{\origttfamily}{}
\let\origttfamily=\ttfamily
\renewcommand{\ttfamily}{\origttfamily \hyphenchar\font=`\-}

\begin{thebibliography}{10}

\bibitem{Bernabei:2008yi}
DAMA, R.~Bernabei et~al., Eur. Phys. J. \textbf{C56} (2008), 333,
  \texttt{0804.2741}.

\bibitem{Ahmed:2008eu}
CDMS, Z.~Ahmed et~al., Phys. Rev. Lett. \textbf{102} (2009), 011301,
  \texttt{0802.3530}.

\bibitem{Angle:2007uj}
XENON, J.~Angle et~al., Phys. Rev. Lett. \textbf{100} (2008), 021303,
  \texttt{0706.0039}.

\bibitem{Fairbairn:2008gz}
M.~Fairbairn and T.~Schwetz, JCAP \textbf{0901} (2009), 037,
  \texttt{0808.0704}.

\bibitem{Savage:2008er}
C.~Savage, G.~Gelmini, P.~Gondolo, and K.~Freese, JCAP \textbf{0904} (2009),
  010, \texttt{0808.3607}.

\bibitem{Foot:2003iv}
R.~Foot, Phys. Rev. \textbf{D69} (2004), 036001, \texttt{hep-ph/0308254}.

\bibitem{Foot:2008nw}
R.~Foot, Phys. Rev. \textbf{D78} (2008), 043529, \texttt{0804.4518}.

\bibitem{Masso:2009mu}
E.~Masso, S.~Mohanty, and S.~Rao, \texttt{0906.1979}.

\bibitem{Feng:2009mn}
J.~L. Feng, M.~Kaplinghat, H.~Tu, and H.-B. Yu, \texttt{0905.3039}.

\bibitem{Khlopov:2008zza}
M.~Y. Khlopov, A.~G. Mayorov, and E.~Y. Soldatov, In *Bled 2008, What comes
  beyond the standard models* 24- 43.

\bibitem{Savage:2004fn}
C.~Savage, P.~Gondolo, and K.~Freese, Phys. Rev. \textbf{D70} (2004), 123513,
  \texttt{astro-ph/0408346}.

\bibitem{Bernabei:2007gr}
R.~Bernabei et~al., Phys. Rev. \textbf{D77} (2008), 023506, \texttt{0712.0562}.

\bibitem{Kopp:2009et}
J.~Kopp, V.~Niro, T.~Schwetz, and J.~Zupan, \texttt{0907.3159}.

\bibitem{TuckerSmith:2001hy}
D.~Tucker-Smith and N.~Weiner, Phys. Rev. \textbf{D64} (2001), 043502,
  \texttt{hep-ph/0101138}.

\bibitem{ArkaniHamed:2000bq}
N.~Arkani-Hamed, L.~J. Hall, H.~Murayama, D.~Tucker-Smith, and N.~Weiner, Phys.
  Rev. \textbf{D64} (2001), 115011, \texttt{hep-ph/0006312}.

\bibitem{Cui:2009xq}
Y.~Cui, D.~E. Morrissey, D.~Poland, and L.~Randall, \texttt{0901.0557}.

\bibitem{Arina:2009um}
C.~Arina, F.-S. Ling, and M.~H.~G. Tytgat, \texttt{0907.0430}.

\bibitem{Chang:2008gd}
S.~Chang, G.~D. Kribs, D.~Tucker-Smith, and N.~Weiner, \texttt{0807.2250}.

\bibitem{MarchRussell:2008dy}
J.~March-Russell, C.~McCabe, and M.~McCullough, \texttt{0812.1931}.

\bibitem{Angloher:2008jj}
G.~Angloher et~al., \texttt{0809.1829}.

\bibitem{Angloher:2004tr}
G.~Angloher et~al., Astropart. Phys. \textbf{23} (2005), 325,
  \texttt{astro-ph/0408006}.

\bibitem{Lewin:1995rx}
J.~D. Lewin and P.~F. Smith, Astropart. Phys. \textbf{6} (1996), 87.

\bibitem{DeJager:1987qc}
H.~De~Vries, C.~W. De~Jager, and C.~De~Vries, Atom. Data Nucl. Data Tabl.
  \textbf{36} (1987), 495.

\bibitem{Fricke}
G.~Fricke et~al., Atomic Data and Nuclear Data Table \textbf{60} (1995), 177.

\bibitem{Duda:2006uk}
G.~Duda, A.~Kemper, and P.~Gondolo, JCAP \textbf{0704} (2007), 012,
  \texttt{hep-ph/0608035}.

\bibitem{Drukier:1986tm}
A.~K. Drukier, K.~Freese, and D.~N. Spergel, Phys. Rev. \textbf{D33} (1986),
  3495.

\bibitem{Smith:2006ym}
M.~C. Smith et~al., Mon. Not. Roy. Astron. Soc. \textbf{379} (2007), 755,
  \texttt{astro-ph/0611671}.

\bibitem{Gelmini:2000dm}
G.~Gelmini and P.~Gondolo, Phys. Rev. \textbf{D64} (2001), 023504,
  \texttt{hep-ph/0012315}.

\bibitem{Dehnen:1997cq}
W.~Dehnen and J.~Binney, Mon. Not. Roy. Astron. Soc. \textbf{298} (1998), 387,
  \texttt{astro-ph/9710077}.

\bibitem{Binney:2008aa}
J.~Binney and S.~Tremaine, Princeton University Press, second ed., 2008.

\bibitem{Bernabei:1996vj}
R.~Bernabei et~al., Phys. Lett. \textbf{B389} (1996), 757.

\bibitem{Bernabei:2007hw}
R.~Bernabei et~al., Eur. Phys. J. \textbf{C53} (2008), 205, \texttt{0710.0288}.

\bibitem{Bernabei:2008yh}
DAMA, R.~Bernabei et~al., Nucl. Instrum. Meth. \textbf{A592} (2008), 297,
  \texttt{0804.2738}.

\bibitem{Bernabei:1998rv}
DAMA, R.~Bernabei et~al., INFN-AE-98-23.

\bibitem{Lebedenko:2008gb}
V.~N. Lebedenko et~al., \texttt{0812.1150}.

\bibitem{Akerib:2003px}
CDMS, D.~S. Akerib et~al., Phys. Rev. \textbf{D68} (2003), 082002,
  \texttt{hep-ex/0306001}.

\bibitem{Angloher:2002in}
G.~Angloher et~al., Astropart. Phys. \textbf{18} (2002), 43.

\bibitem{Aalseth:2008rx}
CoGeNT, C.~E. Aalseth et~al., Phys. Rev. Lett. \textbf{101} (2008), 251301,
  \texttt{0807.0879}.

\bibitem{Lin:2007ka}
TEXONO, S.~T. Lin et~al., Phys. Rev. \textbf{D79} (2009), 061101,
  \texttt{0712.1645}.

\bibitem{Yellin:2002xd}
S.~Yellin, Phys. Rev. \textbf{D66} (2002), 032005, \texttt{physics/0203002}.

\bibitem{Green:2001xy}
A.~M. Green, Phys. Rev. \textbf{D65} (2002), 023520, \texttt{astro-ph/0106555}.

\bibitem{Sorensen:2008ec}
P.~Sorensen et~al., Nucl. Instrum. Meth. \textbf{A601} (2009), 339,
  \texttt{0807.0459}.

\bibitem{Aprile:2008rc}
E.~Aprile et~al., \texttt{0810.0274}.

\bibitem{cdmspc}
{M.~Dragowsky}, private communication.

\bibitem{cresstpc}
{F.~Pr\"obst}, private communication.

\bibitem{cogentpc}
{J.~I.~Collar}, private communication.

\bibitem{texonopc}
{H.~Wong}, private communication.

\bibitem{Nussinov:2009ft}
S.~Nussinov, L.~T. Wang, and I.~Yavin, \texttt{0905.1333}.

\bibitem{Menon:2009qj}
A.~Menon, R.~Morris, A.~Pierce, and N.~Weiner, \texttt{0905.1847}.

\bibitem{iquench}
http://people.roma2.infn.it/~dama/web/nai{\_}que.html.

\bibitem{Ninkovic:2006xy}
J.~Ninkovic et~al., Nucl. Instrum. Meth. \textbf{A564} (2006), 567,
  \texttt{astro-ph/0604094}.

\end{thebibliography}
\bibliographystyle{NewArXiv}

\end{document}